\documentclass[10pt,journal]{IEEEtran}
\IEEEoverridecommandlockouts
\usepackage[linesnumbered,ruled,vlined]{algorithm2e}
\usepackage[noend]{algpseudocode}
\usepackage{amsmath,amssymb,amsfonts}
\usepackage{amsthm}
\usepackage{array}
\usepackage{bm}
\usepackage{booktabs}
\usepackage{cite}
\usepackage{colortbl}
\usepackage{color}
\usepackage{diagbox}
\usepackage{enumitem}
\usepackage{float}
\usepackage{graphicx}
\usepackage{lettrine}
\usepackage{makecell}
\usepackage{mathrsfs}
\usepackage{microtype}
\usepackage{multirow}
\usepackage{multicol}
\usepackage{pifont}
\usepackage{setspace}
\usepackage[caption=false,font=footnotesize]{subfig}
\usepackage{textcomp}
\usepackage{threeparttable}
\usepackage{tikz,xcolor,hyperref}
\usepackage{ulem}
\usepackage{url}
\usepackage{xr-hyper}
\usepackage{zlmtt}

\SetCommentSty{mycommfont}

\SetKwInput{KwInput}{Input}                % Set the Input
\SetKwInput{KwOutput}{Output}              % set the Output
\SetKwInput{KwInitially}{Initially}

\definecolor{colorwheel}{HTML}{000000}
\definecolor{brickred}{HTML}{000000}
\definecolor{mediumorchid}{HTML}{000000}
\definecolor{applegreen}{HTML}{000000}

\def\BibTeX{{\rm B\kern-.05em{\sc i\kern-.025em b}\kern-.08em
    T\kern-.1667em\lower.7ex\hbox{E}\kern-.125emX}}

\newtheorem{theorem}{Theorem}
\newtheorem{lemma}{Lemma}

\newtheorem{Def}{Definition}

\definecolor{lime}{HTML}{A6CE39}

\hypersetup{hidelinks} 
\ifdefined\SHAREAppendixOnly
\fi

\begin{document}

\title{SHARE: A TEE-Assisted Multi-PCH Architecture for Secure and Scalable Off-Chain Payment Services}

\author{
	Lingxiao~Yang,~\IEEEmembership{Member,~IEEE, }
	Xuewen~Dong,~\IEEEmembership{Member,~IEEE, }
	Wei~Wang,~\IEEEmembership{Senior Member,~IEEE, }
	Yong~Yu,~\IEEEmembership{Fellow,~IEEE, }
	Sheng~Gao,~\IEEEmembership{Member,~IEEE, }
	Qiang~Qu,
	and Yulong~Shen,~\IEEEmembership{Senior Member,~IEEE} 
	%and~Jane~Doe,~\IEEEmembership{Life~Fellow,~IEEE}% <-this % stops a space
	\IEEEcompsocitemizethanks{
		\IEEEcompsocthanksitem Lingxiao Yang and Yong Yu are with the School of Artificial Intelligence and Computer Science, Shaanxi Normal University, Xi'an 710119, China (e-mail: lxyang@snnu.edu.cn; yuyong@snnu.edu.cn).
		\IEEEcompsocthanksitem Xuewen Dong is with the School of Computer Science and Technology, Xidian University, the Engineering Research Center of Blockchain Technology Application and Evaluation, Ministry of Education, and also with the Shaanxi Key Laboratory of Blockchain and Secure Computing, Xi'an 710126, China (e-mail: xwdong@xidian.edu.cn).
		\IEEEcompsocthanksitem Wei Wang is with the Beijing Key Laboratory of Security and Privacy in Intelligent Transportation, Beijing Jiaotong University, Beijing 100044, China, and also with the Ministry of Education Key Lab for Intelligent Networks and Network Security, aka MOE KLINNS Lab, Xi'an Jiaotong University, Xi'an 710049, China (e-mail: wangwei1@bjtu.edu.cn).
		\IEEEcompsocthanksitem Sheng Gao is with the School of Information, Central University of Finance and Economics, Beijing 100081, China (e-mail: sgao@cufe.edu.cn).
		\IEEEcompsocthanksitem Qiang Qu is with the Shenzhen Institute of Advanced Technology, Chinese Academy of Sciences, and also with the Huawei Blockchain Lab, Huawei Cloud Tech Co., Ltd, Shenzhen 518055, China (e-mail: qiang@siat.ac.cn).
		\IEEEcompsocthanksitem Yulong Shen is with the School of Computer Science and Technology, Xidian University, and also with the Shaanxi Key Laboratory of Network and System Security, Xi'an 710126, China (e-mail: ylshen@mail.xidian.edu.cn).
		\IEEEcompsocthanksitem \textit{Corresponding authors: Yong Yu, Xuewen Dong.}
		}%\fi% <-this % stops a space
	%\thanks{Manuscript received April 19, 2020; revised August 26, 2020.}
	\thanks{The conference version of this article was presented in part at the IEEE 43rd International Conference on Distributed Computing Systems \cite{yang2023optimal}.}
}

\maketitle
\begin{abstract}
Payment channel hubs (PCHs) improve the reachability of payment channel networks (PCNs) by coordinating off-chain payments between users without direct channels. However, existing hub-based services are commonly constrained by single-hub coordination or sender-computed routes, which limits elasticity under bursty workloads and creates bottlenecks in load balancing, throughput, and privacy protection. This paper presents \textbf{SHARE}, a TEE-assisted multi-PCH architecture for secure and scalable off-chain payment services. SHARE combines cost-aware smooth-node allocation with confidential rate-controlled multipath routing: it solves PCH deployment and client assignment exactly for small networks, applies a randomized double-greedy heuristic for large networks, and uses prior global state together with local requests to spread traffic while preserving channel liquidity. To protect sensitive routing information, SHARE employs TEE-assisted execution and threshold key management, enabling multiple smooth nodes to coordinate concurrent inter-PCH channels without reconstructing complete private keys. A shared-hashlock rule further provides receiver-side payment atomicity, ensuring that an honest receiver accepts only a complete transaction-unit set. We formalize SHARE with a UC-inspired ideal functionality and prove routing confidentiality, execution integrity, and receiver-side acceptance atomicity under explicit leakage and threshold-corruption assumptions. MATLAB simulations and an LND/SGX prototype show that SHARE improves the average transaction success ratio by 43.6\% as transaction size varies and increases normalized throughput by 181.5\% over the evaluated baselines. These results demonstrate that secure multi-hub coordination can turn PCHs from isolated relays into scalable service infrastructure for privacy-preserving off-chain payments.
\end{abstract}


\section{Introduction}\label{sec:introduction}

\lettrine[lines=2]{D}{ecentralized} finance (DeFi) continues to expand, yet the underlying blockchain layer still struggles to deliver service-grade scalability. Since every on-chain transaction requires consensus validation, end-to-end latency can hinder interactive payment services. Off-chain payment channels mitigate this bottleneck by moving frequent transactions outside the consensus path while keeping channel opening, closure, and dispute resolution on-chain \cite{LN, RD}. From a service-computing perspective, they offer low-latency, high-throughput payment services with blockchain-backed settlement finality.

As payment channels interconnect, they form a payment channel network (PCN) that routes off-chain payments through intermediaries when no direct channel exists. This reachability turns payment forwarding into a service orchestration problem: requests must be assigned, traffic must be distributed, and latency and reliability must be maintained under bursty demand. TumbleBit \cite{Ethan2017TumbleBit} introduced an unlinkable payment hub for fast off-chain payments, motivating the payment channel hub (PCH) model.

In the PCH model, participants maintain channels with a hub that coordinates transfers and collects service fees. While this simplifies transaction coordination, it also concentrates service work at one node, limiting flexibility under high-frequency workloads and creating a potential bottleneck. As always-on payment services grow, PCH architectures must support higher concurrency, balanced workloads, and privacy across multiple cooperating hubs.

\begin{figure}[t]
	\centering
	\subfloat[Growing network scale]{
		\begin{minipage}[t]{0.52\linewidth}
			\centering
			\includegraphics[width=1.85in]{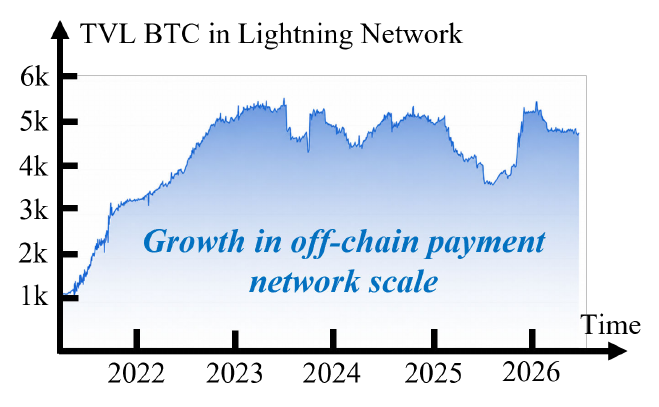}
			\label{Fig1a}
		\end{minipage}%
	}%
	\subfloat[Skewed node load]{
		\begin{minipage}[t]{0.47\linewidth}
			\centering
			\includegraphics[width=1.6in]{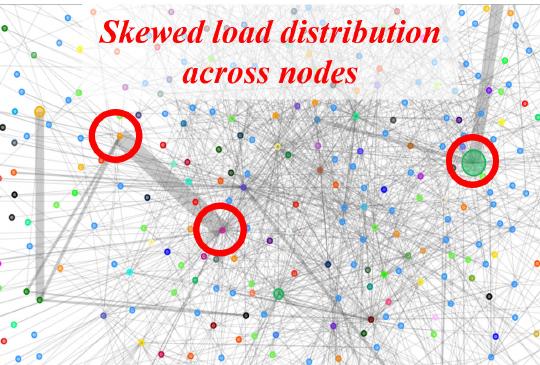}
			\label{Fig1b}
		\end{minipage}
	}%
	\centering
	\caption{Scaling trends and load imbalance in PCNs.}
	\label{Fig1ab}
\end{figure}

\textbf{Motivation.} As shown in Fig. \ref{Fig1a}, the total value locked (TVL) in the Lightning Network \cite{LN} has increased substantially despite short-term fluctuations\footnote{Source: https://defillama.com/protocol/lightning-network}. Although TVL does not directly measure transaction demand, it reflects the growing economic scale of off-chain payment services. Fig. \ref{Fig1b} illustrates a complementary risk: payment traffic may concentrate on a subset of nodes and channels, producing uneven routing load. Together, these observations expose two coupled service-computing challenges: scaling with demand and preventing load concentration. Future off-chain payment services must distribute requests, absorb workload bursts, and maintain responsive forwarding when popular nodes or channels become congested. Three limitations remain.

(i) PCH-based solutions usually rely on one hub to relay payments and hide transactional relationships \cite{Ethan2017TumbleBit, Tairi, qin2023blindhub, Dziembowski2019Perun, Rami2018CommitChains, ge2023accio}, concentrating request processing at a single service point. Multiple hubs alone are insufficient because poor request allocation can still leave some hubs overloaded and others underused. (ii) Collaborative hubs require routing that coordinates concurrent payments without relying only on sender-selected paths. Existing PCN routing protocols include source routing and decentralized schemes based on landmarks, helpers, or embeddings \cite{LN, RD, Flare2016, 0001BBKM19, KhalilG17, Sivaraman2020HighTC, WangXJW19, 10321714, MalavoltaMKM17, RoosMKG18, panwar2024sprite}, but they do not directly address privacy-preserving load balancing across PCHs. (iii) TEE-based approaches protect off-chain confidentiality \cite{LindEPS16, LindNEKSP19, Liao2021, lee2020routee, wang2022sortee}, yet they either require all participants to support TEEs or rely on a single TEE-enabled PCH. Existing approaches therefore do not jointly provide scalable request allocation, load-aware multi-hub routing, and privacy-preserving coordination without a centralized point of failure.

\textbf{Challenges.} SHARE addresses three coupled challenges. (i) \textit{PCH selection and request allocation.} A service architecture must choose suitable hubs, decide how many to activate, and assign clients as network conditions change. (ii) \textit{Multi-PCH routing.} Collaborative hubs must coordinate concurrent payments while respecting liquidity, avoiding contention, and supporting large-value transfers. (iii) \textit{Privacy-preserving collaborative routing.} Routing across multiple service nodes must protect payment amounts, endpoint identities, and transaction relationships.

\textbf{Contributions.} This paper presents \textbf{SHARE} (\textbf{\underline{S}}ecure \textbf{\underline{H}}ub \textbf{\underline{A}}llocation and \textbf{\underline{R}}outing \textbf{\underline{E}}fficiency), a TEE-assisted multi-PCH architecture for secure and scalable off-chain payment services. Following Splicer \cite{yang2023optimal}, we call the selected PCHs \textit{smooth nodes}; in SHARE, each smooth node is TEE-enabled. SHARE distributes payment processing and routing across multiple smooth nodes to balance network load and avoid single-hub bottlenecks. The main contributions are as follows.

\begin{itemize}[leftmargin=*]
\item[$\bullet$] \textbf{Cost-aware multi-PCH allocation.} We formulate smooth-node deployment and client assignment as an NP-hard optimization problem that captures deployment, client-management, and inter-smooth-node synchronization costs. SHARE solves small instances exactly through MILP and large instances through a randomized double-greedy heuristic based on supermodular structure.

\begin{table*}[!tbp]
	\centering
	\caption{Comparison of representative PCN routing, PCH, and TEE-assisted schemes.}
	\label{tab-routing-compare}
	\scriptsize
	\setlength{\tabcolsep}{2.1pt}
	\renewcommand{\arraystretch}{0.92}
	
	\begin{threeparttable}
		\begin{tabular*}{\textwidth}{
				@{\extracolsep{\fill}} l c c c c c c c @{}
			}
			\toprule
			Schemes &
			\makecell[c]{TEE\\locus} &
			\makecell[c]{Multi-path\\routing} &
			\makecell[c]{Multiple\\service nodes} &
			\makecell[c]{Load-aware\\coord.} &
			\makecell[c]{Value\\privacy} &
			\makecell[c]{Endpoint / rel.\\privacy} &
			\makecell[c]{Formal\\proof} \\
			\midrule
			
			TumbleBit, A$^{2}$L \cite{Ethan2017TumbleBit,Tairi}
			& None & \ding{55} & \ding{55} & N/A & \ding{55} & \ding{51} & \ding{51} \\
			
			BlindHub, Accio \cite{qin2023blindhub,ge2023accio}
			& None & \ding{55} & \ding{55} & N/A & \ding{51} & \ding{51} & \ding{51} \\
			
			SilentWhispers \cite{MalavoltaMKM17}
			& None & \ding{51} & N/A & -- & \ding{51} & \ding{51} & \ding{51} \\
			
			SpeedyMurmurs \cite{RoosMKG18}
			& None & \ding{51} & N/A & -- & \ding{51} & \ding{51} & -- \\
			
			Flare \cite{Flare2016}
			& None & \ding{51} & N/A & -- & -- & -- & -- \\
			
			Flash \cite{WangXJW19}
			& None & \ding{51} & N/A & \ding{51} & -- & -- & -- \\
			
			Sprite \cite{panwar2024sprite}
			& None & \ding{51} & N/A & -- & \ding{51} & \ding{51} & \ding{51} \\
			
			Teechain, Speedster \cite{LindNEKSP19,Liao2021}
			& Clients & -- & N/A & N/A & \ding{51} & -- & -- \\
			
			Twilight \cite{dotan2022twilight}
			& Relays & -- & \ding{51} & -- & \ding{51} & \ding{51} & \ding{51} \\
			
			RouTEE \cite{lee2020routee}
			& PCH & \ding{55} & \ding{55} & -- & \ding{51} & -- & -- \\
			
			\textsc{SorTEE} \cite{wang2022sortee}
			& Service nodes & \ding{51} & \ding{51} & -- & \ding{51} & \ding{51} & -- \\
			
			Splicer \cite{yang2023optimal}
			& None & \ding{51} & \ding{51} & \ding{51} & -- & \ding{51} & -- \\
			
			\textbf{SHARE}
			& PCHs & \ding{51} & \ding{51} & \ding{51}
			& \ding{51} & \ding{51}\tnote{a} & \ding{51} \\
			
			\bottomrule
		\end{tabular*}
		
		\begin{tablenotes}[flushleft]
			\scriptsize
			\item[] Notes: ``\ding{51}'': explicitly supported;
			``\ding{55}'': not supported;
			``N/A'': not applicable;
			``--'': not explicitly established or not a main focus in the cited paper.
			Grouped schemes in one row share the listed feature values. $^a$ Relationship unlinkability is inherited from the underlying PCH construction and is not reproved in this work.
		\end{tablenotes}
	\end{threeparttable}
\end{table*}

\item[$\bullet$] \textbf{TEE-assisted distributed routing.} SHARE combines previous-epoch global state with local payment requests to coordinate multipath rates, apply congestion control, balance load, and preserve channel liquidity. A receiver-generated shared hashlock ensures that honest receivers accept only complete transaction-unit sets, while attested state management supports concurrent logical channels between smooth nodes.

\item[$\bullet$] \textbf{Formal security analysis.} We define a UC-inspired ideal functionality and give a property-based proof showing routing confidentiality, execution integrity, and receiver-side acceptance atomicity under the stated leakage model and threshold-corruption bound. Relationship unlinkability is inherited from the underlying PCH construction and is not reproved here.

\item[$\bullet$] \textbf{Prototype implementation and evaluation.} MATLAB simulations and an LND/SGX prototype \cite{Anati2013,lnd} show that SHARE improves average transaction success ratio by 43.6\% as transaction size varies and normalized throughput by 181.5\% over the comparison set.
\end{itemize}

\iffalse
\noindent\textbf{Outline.} The rest of this paper proceeds as follows. Section \ref{section:Relatedwork} reviews related work. Section \ref{pre} presents some preliminaries about PCNs. Section \ref{sec_Problem} states the problems in this work. In Section \ref{sec_SD}, we present the optimization PCHs placement problem solutions and the design of the routing protocol. We show the experimental results in Section \ref{Section:Evaluation}. Finally, Section \ref{conclu} concludes the paper.
\fi

\section{Related Work}\label{section:Relatedwork}
This section reviews prior work from three perspectives: PCH mechanisms for off-chain payment services, PCN routing strategies for service coordination, and TEE-assisted approaches to privacy-preserving off-chain execution.

\textbf{PCH mechanisms.} TumbleBit \cite{Ethan2017TumbleBit} introduced a Bitcoin-compatible untrusted payment hub that supports fast off-chain payments while preventing the intermediary from linking payers to payees. A$^{2}$L \cite{Tairi} uses anonymous atomic locks to provide a backward-compatible PCH protocol with atomicity and payment unlinkability. BlindHub \cite{qin2023blindhub} introduces blind adaptor signatures and flexible blind conditional signatures to support variable-amount payments while preserving atomicity, relationship anonymity, and value privacy. Accio \cite{ge2023accio} combines randomizable signatures with updatable commitments to support variable-amount payments without the overhead of non-interactive zero-knowledge proofs. These schemes strengthen private hub-mediated payments, but they retain a single-hub architecture and do not address request allocation or collaborative routing across multiple PCHs.

\textbf{Routing strategies.} Most PCH protocols focus on hub-mediated transfers rather than general multi-hop routing. PCN routing remains relevant because it determines how payment requests are discovered, split, and forwarded under changing liquidity and network load. Beyond sender-computed source routing, decentralized protocols reduce the sender's dependence on complete topology information. SilentWhispers \cite{MalavoltaMKM17} adapts landmark-based routing to support privacy-preserving transactions in decentralized credit networks. SpeedyMurmurs \cite{RoosMKG18} uses embedding-based path discovery and on-demand stabilization to improve routing efficiency under topology changes. Flare \cite{Flare2016} and Flash \cite{WangXJW19} explore route discovery with partial topology knowledge and liquidity-aware path selection, while Sprite \cite{panwar2024sprite} separates offline path discovery from online transaction processing to support concurrent private payments with lower message complexity. These protocols improve decentralized path discovery, but they are not designed to jointly optimize multi-PCH allocation, inter-hub load balancing, and confidential coordination among hubs.

\textbf{TEE-based solutions.} Teechain \cite{LindNEKSP19} uses TEEs to execute off-chain transactions asynchronously with respect to the underlying blockchain. Speedster \cite{Liao2021} develops an account-based state-channel system in which enclave-backed certified channels support dispute-free off-chain operation and multiparty contracts. Twilight \cite{dotan2022twilight} uses TEE-enabled relays to enforce a noisy relay mechanism for differential privacy, while \textsc{SorTEE} \cite{wang2022sortee} delegates path computation to TEE-enabled service nodes to reduce user-side routing overhead and protect payment information. RouTEE \cite{lee2020routee} implements a TEE-protected routing hub that conceals payment information, improves liquidity utilization, and removes the need for users to continuously monitor the blockchain. Despite these advances, existing TEE-assisted systems either require broad client-side TEE deployment, depend on a single TEE-enabled PCH, or do not jointly address cost-aware multi-PCH allocation and load-aware confidential routing.

\textbf{Comparison.} Table \ref{tab-routing-compare} summarizes representative schemes along dimensions relevant to SHARE. Conventional PCH protocols provide private hub-mediated payments but funnel requests through one intermediary. Decentralized routing protocols reduce sender-side topology knowledge and computation, but they do not coordinate multiple service hubs under concurrent workloads. TEE-assisted systems protect sensitive off-chain execution, yet they do not combine selective hub-side TEE deployment with cost-aware allocation and load-aware inter-hub routing. SHARE targets this gap by deploying TEEs only at selected PCHs and distributing routing computation across them. Our previous work, Splicer \cite{yang2023optimal}, establishes a multi-PCH architecture with cost-aware hub allocation and rate-controlled routing. SHARE extends this foundation with TEE-protected routing computation, concurrent inter-PCH channel management, and a UC-inspired ideal-functionality specification with a scoped property-based security analysis. Relationship unlinkability is inherited from the underlying PCH construction and is not reproved in this work.

\section{SHARE: System Overview}\label{sec_Problem}
This section presents the system model, workflow, and threat assumptions of SHARE as an off-chain payment service architecture. We first define the participating entities and their interactions, then describe the payment workflow, and finally state the adversary capabilities and security objectives that guide the protocol design and security proof.

\subsection{System Model}\label{sec_SM}
\textbf{Entities.} SHARE involves two types of entities:

\textbf{\textit{Client.}} Clients are lightweight end users in the PCN who initiate or receive payments. They do not execute the routing protocol themselves. Instead, each client delegates payment coordination to one assigned smooth node and communicates only with that smooth node through a secure channel.

\textbf{\textit{Smooth node.}} A smooth node is a TEE-enabled hub that executes the routing logic inside an enclave. Smooth nodes collaboratively process client payment requests, determine route splitting and forwarding decisions, and generate execution evidence for later verification. A subset of $\iota$ smooth nodes forms the key management group (KMG) and runs a robust threshold public-key encryption scheme $\mathcal{TE}=(\textsf{DKG},\textsf{Enc},\textsf{PDec},\textsf{VerifyShare},\textsf{Combine})$ \cite{Boneh2006Threshold}. Let $t$ be the maximum number of statically corrupted KMG members. SHARE sets the decryption threshold to $\tau_{\rm dec}=t+1$ and requires $\iota\geq 2t+1$. During offline preprocessing, the KMG repeatedly runs distributed key generation \cite{GRJ1999} to fill a pool of unused one-time entries. Each entry contains public and verification material and one sealed private-key share in every KMG enclave. Online, the KMG atomically binds one unused entry to a fresh transaction or TU identifier $x$, exposes $(\textsf{pk}_x,\textsf{vk}_x)$, and marks the entry as allocated. Every private-key share $\textsf{sk}_x^{(q)}$ remains sealed inside its holder's enclave; no complete private key is materialized, exported, or reconstructed.

\textbf{Workflow.} A PCN is modeled as a graph $\mathbb{G}=(\mathbb{V}, \mathbb{E})$, where $\mathbb{V}$ is the set of nodes and $\mathbb{E}$ is the set of payment channels. Let $\mathbb{V}_{\rm CLI} \subseteq \mathbb{V}$ denote the set of clients and let $\mathbb{V}_{\rm SN} \subseteq \mathbb{V}$ denote the set of smooth nodes, such that $\mathbb{V}_{\rm CLI}=\mathbb{V}\setminus \mathbb{V}_{\rm SN}$. We further define an assignment function
\begin{equation}
a: \mathbb{V}_{\rm CLI} \rightarrow \mathbb{V}_{\rm SN},
\label{eq:assignment-function}
\end{equation}
where $a(P)$ returns the smooth node assigned to client $P$. In the default deployment, each client maintains a direct channel to exactly one assigned smooth node, while smooth nodes may communicate with each other through the hub network and the KMG.

\begin{figure}[t]
	\centering
	\includegraphics[width=\columnwidth]{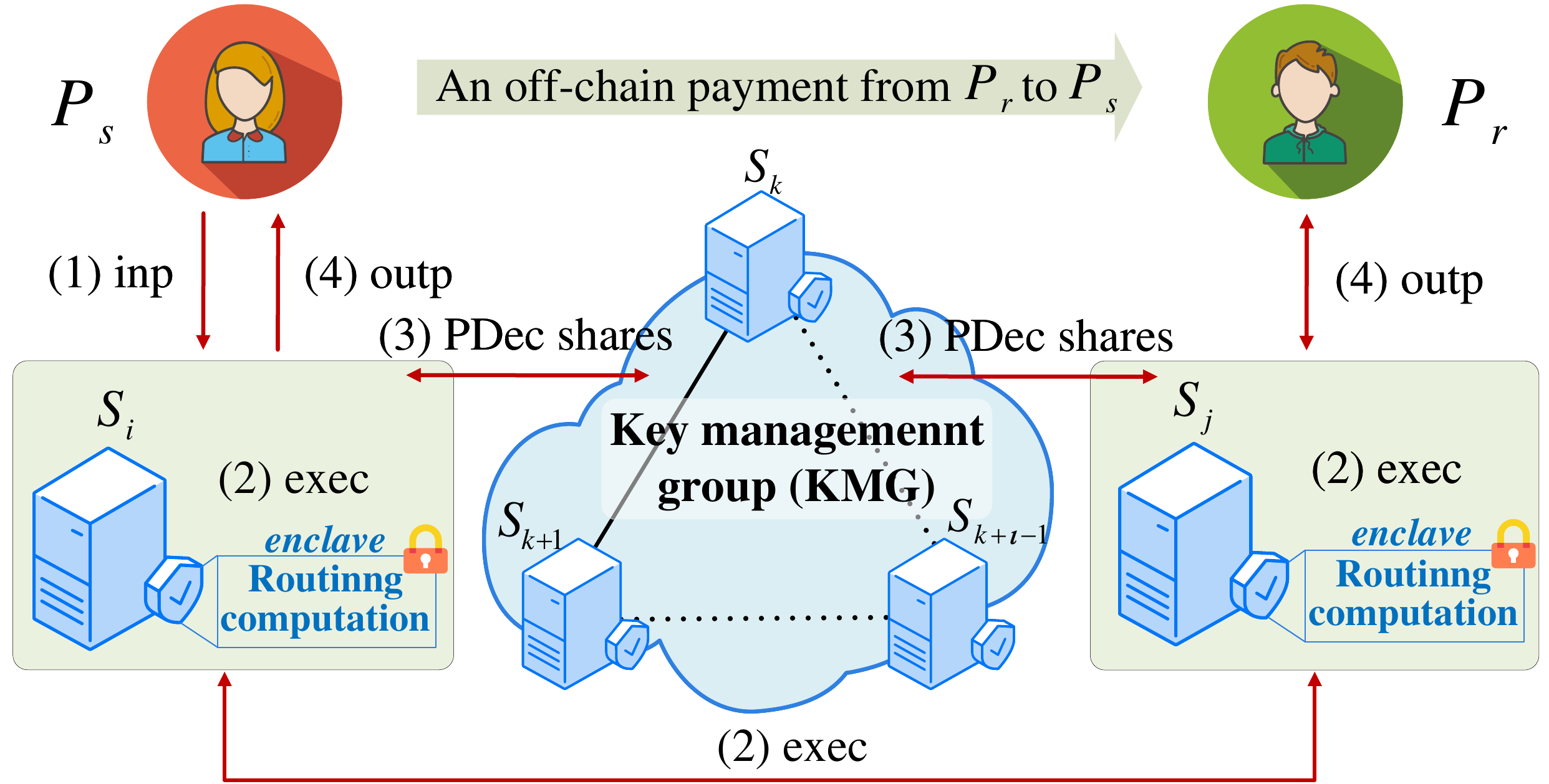}
	\caption{System model of SHARE.}
	\label{workflow}
\end{figure}

As shown in Fig. \ref{workflow}, when a client $P_s \in \mathbb{V}_{\rm CLI}$ initiates a payment to another client $P_r \in \mathbb{V}_{\rm CLI}$, the assigned smooth nodes are $S_i=a(P_s)$ and $S_j=a(P_r)$, respectively. The payment workflow consists of three phases: \textit{initialization}, \textit{processing}, and \textit{acknowledgment}.

\textit{a) Payment initialization phase:} For simplicity, we omit the channel-opening procedure and assume that all channels already contain sufficient initial deposits, following prior work \cite{LiMZ20, Ge2022ShadufNP}. Before initiating a payment, $P_r$ samples a one-time payment preimage $z_{\textsf{tid}}\leftarrow\$\{0,1\}^{\lambda}$, sends the invoice commitment $h_{\textsf{tid}}=\textsf{H}(z_{\textsf{tid}})$ to $P_s$, and retains $z_{\textsf{tid}}$. During setup, $S_i$ installs $\textsf{prog}$ in $\mathcal{G}_{\rm att}$ and obtains an enclave identifier $\textsf{eid}_i$, an attestation public key $\textsf{mpk}_{\rm att}$, and platform evidence binding $(\textsf{eid}_i,\textsf{prog},\textsf{mpk}_{\rm att})$. The corresponding signing key remains inaccessible inside $\mathcal{G}_{\rm att}$. The client $P_s$ verifies this evidence and establishes an attestation-authenticated secure channel with the enclave of $S_i$; $P_r$ does the same with the enclave of $S_j$. After the secure channels are established, $P_s$ sends $\textsf{pay}_{\rm req}$ and $h_{\textsf{tid}}$ to $S_i$. The enclave samples a fresh transaction identifier $\textsf{tid}\leftarrow\$\{0,1\}^{\lambda}$ and asks the KMG to allocate an unused pre-generated threshold-key entry. The KMG atomically binds the entry to $\textsf{tid}$ and returns only $(\textsf{pk}_{\textsf{tid}},\textsf{vk}_{\textsf{tid}})$; the private-key shares $\{\textsf{sk}_{\textsf{tid}}^{(q)}\}_{q\in[\iota]}$ remain sealed in the corresponding KMG enclaves. Smooth node $S_i$ sends $(\textsf{tid},\textsf{pk}_{\textsf{tid}})$ to $P_s$ and initializes $\textsf{state}_{\textsf{tid}}=(\textsf{tid},\theta_{\textsf{tid}},h_{\textsf{tid}})$, where $\theta_{\textsf{tid}}$ indicates whether the transaction has completed.

\textit{b) Payment processing and acknowledgment phases:} Fig. \ref{workflow} illustrates the payment execution process, which consists of a processing stage followed by an acknowledgment stage.

\textbf{Step (1):} The transaction begins when $P_s$ constructs the payment request
\begin{equation}
D_{\textsf{tid}}=(P_s, P_r, \textsf{val}_{\textsf{tid}},h_{\textsf{tid}}),
\label{eq:payment-request}
\end{equation}
where $\textsf{val}_{\textsf{tid}}$ is the payment amount and $h_{\textsf{tid}}$ is the receiver-generated payment hash. $P_s$ then computes $\textsf{inp}=\mathcal{TE}.\textsf{Enc}(\textsf{pk}_{\textsf{tid}},D_{\textsf{tid}})$ and sends $(\textsf{tid},\textsf{inp})$ together with a conditional transfer locked by $h_{\textsf{tid}}$ to $S_i$.

\textbf{Step (2--3):} Smooth node $S_i$ loads $\textsf{inp}$ into its enclave. The enclave produces authorization evidence $\pi_{\textsf{tid}}^{\rm dec}$ that binds $(\textsf{eid}_i,\textsf{prog},\textsf{tid},\textsf{H}(\textsf{inp}))$. A KMG member responds only after verifying this evidence, confirming that $\textsf{tid}$ is bound to an allocated entry, and establishing an attestation-authenticated channel terminating at $\textsf{eid}_i$. Each responsive member $q$ then returns $\delta_{\textsf{tid}}^{(q)}=\mathcal{TE}.\textsf{PDec}(\textsf{pk}_{\textsf{tid}},\textsf{sk}_{\textsf{tid}}^{(q)},\textsf{inp})$ directly to the enclave. The enclave accepts only shares satisfying $\mathcal{TE}.\textsf{VerifyShare}(\textsf{pk}_{\textsf{tid}},\textsf{vk}_{\textsf{tid}},\textsf{inp},\delta_{\textsf{tid}}^{(q)})=1$. After collecting a set $Q$ of at least $\tau_{\rm dec}$ valid shares, it recovers
\begin{equation}
D_{\textsf{tid}}=\mathcal{TE}.\textsf{Combine}(\textsf{pk}_{\textsf{tid}},\textsf{vk}_{\textsf{tid}},\textsf{inp},\{\delta_{\textsf{tid}}^{(q)}\}_{q\in Q}).
\label{eq:threshold-decryption}
\end{equation}
If fewer than $\tau_{\rm dec}$ valid shares are available, the transaction aborts. The routing program $\textsf{prog}$ then partitions $D_{\textsf{tid}}$ into $K$ transaction units (TUs) $\{D_{\textsf{tuid}_k}\}_{k=1}^{K}$, where each TU has a unique identifier $\textsf{tuid}_k$. For every TU, $S_i$ creates $\textsf{state}_{\textsf{tuid}_k}=(\textsf{tuid}_k,\theta_{\textsf{tuid}_k})$, where $\theta_{\textsf{tuid}_k}$ records whether that TU has completed successfully and
\begin{equation}
\theta_{\textsf{tid}}=\bigwedge_{k=1}^{K}\theta_{\textsf{tuid}_k}.
\label{eq:completion-flag}
\end{equation}

For each $\textsf{tuid}_k$, the KMG atomically allocates and binds another unused pre-generated entry, then publishes $(\textsf{pk}_{\textsf{tuid}_k},\textsf{vk}_{\textsf{tuid}_k})$. Every TU carries the same payment hash $h_{\textsf{tid}}$ and is forwarded as an HTLC with a path-valid expiry. Smooth node $S_i$ encrypts $D_{\textsf{tuid}_k}$ under $\textsf{pk}_{\textsf{tuid}_k}$ and dispatches the ciphertext and conditional transfer over the selected path without waiting for other TUs to finish. The enclave of $S_j$ obtains authorization evidence binding the TU ciphertext to its attested program, requests shares through its attested enclave channel, verifies them, and combines any $\tau_{\rm dec}$ valid shares to recover the TU; the complete private key is never reconstructed. If the TU is valid, its HTLC is irrevocably locked, and sufficient expiry margin remains, $S_j$ returns $\textsf{READY}_{\textsf{tuid}_k}$ without revealing the preimage. Only after all $K$ readiness messages have arrived does $S_j$ request $z_{\textsf{tid}}$ from $P_r$, verify $\textsf{H}(z_{\textsf{tid}})=h_{\textsf{tid}}$, and initiate fulfillment of the complete HTLC set with the common preimage. In accordance with Basic MPP semantics, the destination does not fulfill an incomplete set and, if it fulfills one TU, it issues fulfillment for the entire set. The preimage then propagates backward along each path under the standard HTLC protocol, and $S_i$ marks a TU complete only after observing its fulfillment. If any TU fails or expires before preimage release, $P_r$ withholds $z_{\textsf{tid}}$; each pending HTLC then follows its timeout branch. Thus, an honest receiver accepts either the complete payment or none of its TUs. After preimage release, individual paths may resolve at different times under adversarial scheduling, so SHARE claims receiver-side acceptance atomicity rather than simultaneous end-to-end settlement. Once all fulfillment acknowledgments are received, $S_i$ updates $\textsf{state}_{\textsf{tid}}$ and asks $\mathcal{G}_{\rm att}$ to finalize the state and generate
\begin{equation}
\sigma_{\textsf{tid}}\leftarrow\mathcal{G}_{\rm att}.\textsf{resume}(\textsf{eid}_i,(\textsf{finalize},\textsf{state}_{\textsf{tid}})),
\label{eq:exec-proof}
\end{equation}
where the signing key remains internal to $\mathcal{G}_{\rm att}$. The signature covers the enclave identity, program measurement, command, and final state and is verified under the previously attested $\textsf{mpk}_{\rm att}$.

\textbf{Step (4):} The release of $z_{\textsf{tid}}$ makes the complete HTLC set claimable and allows $P_r$ to accept the consolidated payment. Each conditional transfer then resolves through the underlying HTLC success path, which may complete at different times. $P_r$ generates $\textsf{ACK}_{\textsf{tid}}$ after accepting the complete set, while $S_i$ finalizes success for $P_s$ only after receiving all TU fulfillment acknowledgments. Because clients verified the platform quote and its binding to $\textsf{mpk}_{\rm att}$ during channel setup, they verify $\sigma_{\textsf{tid}}$ locally under that key; the platform attestation service is not invoked for every transaction. The forwarding fees paid by the initiator incentivize relay nodes along the routing path.

\subsection{Threat Model}\label{privacy}
\textbf{Adversary model.} We consider a static PPT adversary that can observe the public network, corrupt smooth-node hosts, and control their operating systems and network stacks. Before protocol setup, the adversary may corrupt at most $t$ of the $\iota$ KMG members, where $\iota\geq 2t+1$ and $\tau_{\rm dec}=t+1$. It may drop, delay, replay, or reorder messages handled by corrupted hosts and learn any information exposed outside their enclaves. However, it cannot break the confidentiality or integrity of honest TEEs, extract a private-key share from an honest KMG enclave, cause an honest KMG enclave to release a share outside an authorized attested channel, combine at most $t$ shares to recover a plaintext or private key, or forge valid platform evidence or enclave-generated execution proofs.

\textbf{Security objectives.} SHARE is designed to satisfy the following goals:
\begin{itemize}[leftmargin=*]
	\item \textit{Routing confidentiality:} sensitive routing inputs, including payment amounts and transaction payloads, should not be revealed to entities outside the trusted execution environment.
	\item \textit{Conditional relationship unlinkability:} assuming the underlying PCH construction provides relationship unlinkability, splitting a payment must not introduce additional endpoint-linkage leakage beyond the explicitly modeled metadata. This inherited property is not part of SHARE's UC theorem.
	\item \textit{Execution integrity:} the externally visible execution proof should allow a verifier to check that the routing program was executed correctly.
	\item \textit{Receiver-side acceptance atomicity:} for honest endpoints, the receiver should release the shared preimage only for a complete HTLC set and should never accept a partial payment. An incomplete set remains unfulfilled and its pending HTLCs resolve through their timeout branches.
\end{itemize}

\textbf{Assumptions.} We assume that all parties trust Intel processors and the SGX attestation infrastructure, instantiated by IAS in legacy EPID deployments or by DCAP/PCS in ECDSA deployments. KMG members use authenticated secure channels and an authenticated broadcast channel during $\mathcal{TE}.\textsf{DKG}$, and release partial decryptions only to enclaves whose evidence and program measurement have been verified. The threshold encryption scheme is robust and IND-CCA secure against exposure of at most $t$ private-key shares, and every partial decryption is verifiable. Confidentiality holds against the static corruption of at most $t$ KMG members, whereas threshold-decryption availability requires at least $\tau_{\rm dec}$ responsive members. The underlying PCN enforces standard HTLC success and timeout semantics and the Basic MPP terminal rule\footnote{https://github.com/lightning/bolts/blob/master/04-onion-routing.md}: a conditional transfer can be fulfilled only with a valid preimage before expiry; an incomplete HTLC set is not fulfilled; and a terminal node that fulfills one member of a complete set issues fulfillment for the entire set. The receiver releases its payment preimage only after all $K$ TUs are irrevocably locked. These assumptions guarantee atomic acceptance at the honest receiver, but do not impose a bound on network delay or require all paths to resolve simultaneously. The offline pool must contain enough unused entries for the admitted workload; if it is exhausted, new payments wait for a preprocessing refill rather than falling back to online DKG. Clients trust a smooth node only after verifying its platform evidence and establishing the corresponding attested channel.

\textbf{Side-channel mitigation.} Existing TEE implementations remain vulnerable to side-channel attacks \cite{MurdockOGBGP20, fei2021security}. To reduce this risk, SHARE uses side-channel-resistant cryptographic libraries\footnote{https://www.trustedfirmware.org/projects/mbed-tls/}. Side-channel hardening is not the main focus of this paper and is treated as an orthogonal engineering concern.

\begin{figure*}[!tbp]
\centering
\includegraphics[width=7.1in]{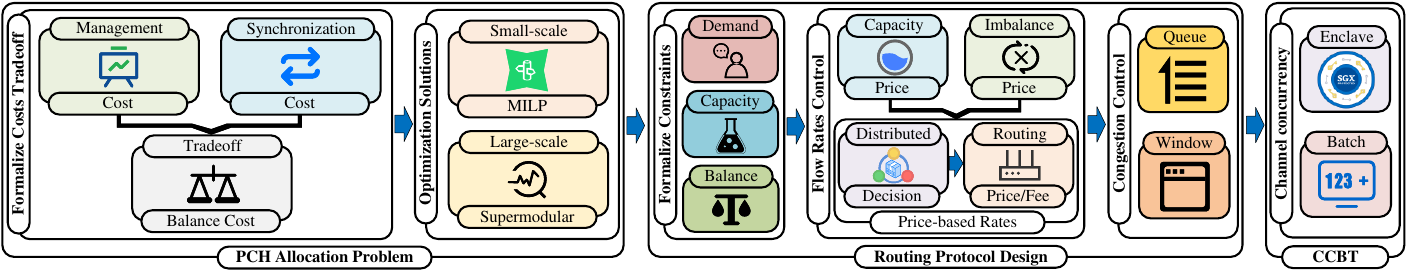}
\caption{The overall design structure of SHARE.}
\label{system-overview}
\end{figure*}

\section{SHARE: System Design}\label{sec_SD}

\subsection{Design Overview}
Fig. \ref{system-overview} summarizes the SHARE design. The system consists of three tightly coupled components. First, we formulate the smooth-node allocation problem, whose goal is to balance deployment and client-management costs against inter-smooth-node synchronization costs. For small networks, we solve the allocation problem exactly as a mixed-integer linear program (MILP). For larger networks, we exploit its supermodular structure under uniform synchronization costs and use a scalable randomized heuristic.

Second, we design a rate-based distributed routing protocol for smooth nodes. The protocol combines global state from the previous epoch with current client requests to compute routing prices, adjust per-path flow rates, and prevent both capacity overload and directional liquidity depletion. This allows SHARE to support concurrent multipath forwarding while keeping channel utilization stable.

Third, we analyze channel concurrency. Smooth nodes execute routing logic inside TEEs and obtain threshold decryption shares from the KMG. SHARE can therefore operate multiple pre-funded logical channels between PCHs without an on-chain transaction for every off-chain state update. We model the concurrency benefit using the Universal Scalability Law (USL) and use batch processing to reduce enclave state-management overhead.

\subsection{Allocation Problem Formulation}\label{detail_problem}
For brevity, we refer readers to Appendix B for the long-term candidate selection mechanism and to Appendix C for the broader problem context. Here we formalize the allocation problem that selects active smooth nodes from the candidate set.

Let $\mathbb{V}_{\rm SNC}$ denote the smooth-node candidate set and let $\mathbb{V}_{\rm CLI}$ denote the client set. We use two binary decision vectors:
\begin{align}
\boldsymbol{x}&=\{x_n \in \{0,1\}: n \in \mathbb{V}_{\rm SNC}\}, \label{eq:deployment-plan}\\
\boldsymbol{y}&=\{y_{mn} \in \{0,1\}: m \in \mathbb{V}_{\rm CLI}, n \in \mathbb{V}_{\rm SNC}\}. \label{eq:assignment-plan}
\end{align}
Here, $x_n=1$ means that candidate $n$ is deployed as an active smooth node, and $y_{mn}=1$ means that client $m$ is assigned to smooth node $n$. Each client is assigned to exactly one active smooth node, and at least one smooth node must be deployed:
\begin{small}
\begin{gather}
\sum_{n \in \mathbb{V}_{\rm SNC}} y_{mn}=1,\quad \forall m \in \mathbb{V}_{\rm CLI}, \label{constraint13}\\
y_{mn} \le x_n,\quad \forall m \in \mathbb{V}_{\rm CLI}, n \in \mathbb{V}_{\rm SNC}, \label{constraint14}\\
\sum_{n \in \mathbb{V}_{\rm SNC}}x_n\geq 1. \label{constraint15}
\end{gather}
\end{small}

The allocation objective balances three costs. The management cost $\mathcal{C}_M(\boldsymbol{y})$ captures client-to-smooth-node assignment overhead, the deployment cost $\mathcal{C}_D(\boldsymbol{x})$ penalizes activating smooth nodes, and the synchronization cost $\mathcal{C}_S(\boldsymbol{x},\boldsymbol{y})$ captures inter-smooth-node coordination overhead:
\begin{small}
\begin{align}
\mathcal{C}_M(\boldsymbol{y})
&=\sum_{m \in \mathbb{V}_{\rm CLI}}\sum_{n \in \mathbb{V}_{\rm SNC}}\zeta_{mn}y_{mn}, \label{cost1}\\
\mathcal{C}_D(\boldsymbol{x})
&=\sum_{n \in \mathbb{V}_{\rm SNC}}f_nx_n, \label{cost-deployment}\\
\mathcal{C}_S(\boldsymbol{x},\boldsymbol{y})
&=\sum_{n \in \mathbb{V}_{\rm SNC}}\sum_{\substack{l \in \mathbb{V}_{\rm SNC}\\l\ne n}}
\left(\delta_{nl}x_l\sum_{m \in \mathbb{V}_{\rm CLI}}y_{mn}
+\frac{1}{2}\epsilon_{nl}x_nx_l\right). \label{cost2}
\end{align}
\end{small}
Here, $\zeta_{mn}\geq0$ is the management cost of assigning client $m$ to smooth node $n$, $f_n\geq0$ is the deployment cost of node $n$, $\delta_{nl}\geq0$ is the per-client synchronization cost from $n$ to $l$, and $\epsilon_{nl}=\epsilon_{ln}\geq0$ is the fixed pairwise synchronization overhead. The factor $1/2$ prevents double counting of the fixed overhead.

We define the balanced objective as
\begin{equation}
\mathcal{C}_B(\boldsymbol{x},\boldsymbol{y})
=\mathcal{C}_M(\boldsymbol{y})+\mathcal{C}_D(\boldsymbol{x})
+\omega\mathcal{C}_S(\boldsymbol{x},\boldsymbol{y}),
\label{eq:balanced-cost}
\end{equation}
where $\omega\geq0$ controls the weight of synchronization overhead. The PCH allocation problem is
\begin{equation}
\min_{\boldsymbol{x},\boldsymbol{y}}\mathcal{C}_B(\boldsymbol{x},\boldsymbol{y})
\quad\text{s.t. Equations }\eqref{constraint13}\text{--}\eqref{constraint15}.
\label{eq:allocation-problem}
\end{equation}

\textbf{NP-hardness.} Set $\omega=0$. Equation \eqref{eq:allocation-problem} then selects facilities with opening costs $f_n$ and assigns every client $m$ to one open facility at cost $\zeta_{mn}$. This is precisely uncapacitated facility location, which is NP-hard. Therefore, the general SHARE allocation problem is NP-hard.

\subsection{Allocation Algorithms}\label{solutions}
\textbf{Small-scale exact solution.} For small networks, we linearize the binary products by introducing
\begin{small}
\begin{gather}
\boldsymbol{\vartheta}=\{\vartheta_{nl}\in\{0,1\}:n,l\in\mathbb{V}_{\rm SNC},\ n\ne l\},\\
\boldsymbol{\varphi}=\{\varphi_{nlm}\in\{0,1\}:n,l\in\mathbb{V}_{\rm SNC},\ n\ne l,\ m\in\mathbb{V}_{\rm CLI}\}.
\end{gather}
\end{small}
The following constraints enforce $\vartheta_{nl}=x_nx_l$ and $\varphi_{nlm}=x_ly_{mn}$:
\begin{small}
\begin{align}
\vartheta_{nl}&\le x_n,& \vartheta_{nl}&\le x_l,&
\vartheta_{nl}&\ge x_n+x_l-1, \label{constraint_v2}\\
\varphi_{nlm}&\le x_l,& \varphi_{nlm}&\le y_{mn},&
\varphi_{nlm}&\ge x_l+y_{mn}-1. \label{constraint_v4}
\end{align}
\end{small}
The exact MILP objective is
\begin{align}
\min\quad
&\sum_{m,n}\zeta_{mn}y_{mn}+\sum_nf_nx_n \notag\\
&+\omega\sum_n\sum_{l\ne n}\left(\delta_{nl}\sum_m\varphi_{nlm}
+\frac{1}{2}\epsilon_{nl}\vartheta_{nl}\right), \label{eq:allocation-milp}
\end{align}
subject to Equations \eqref{constraint13}--\eqref{constraint15}, \eqref{constraint_v2}, and \eqref{constraint_v4}, with indices ranging over their corresponding client and candidate sets. Standard branch-and-bound solvers can obtain the optimum for small instances.

\textbf{Large-scale heuristic.} Once a deployment set is fixed, the best assignment is separable across clients.

\begin{lemma}\label{lenmma1}
Given a nonempty deployment set $\mathcal{X}$, the optimal assignment for each client $m$ is
\begin{align}
n_m^*\in\arg\min_{n\in\mathcal{X}}
\left\{\zeta_{mn}+\omega\sum_{l\in\mathcal{X}\setminus\{n\}}\delta_{nl}\right\}, \label{eq:optimal-assignment}
\end{align}
with $y_{mn_m^*}=1$ and $y_{mn}=0$ for $n\ne n_m^*$.
\end{lemma}

Thus, the main combinatorial decision is the deployment set. Let $\mathcal{S}$ denote the candidate-node ground set and let $\varnothing\ne\mathcal{X}\subseteq\mathcal{S}$. Substituting Equation \eqref{eq:optimal-assignment} into the objective gives $f(\mathcal{X})=\mathcal{C}_B(\boldsymbol{x}_{\mathcal{X}},\boldsymbol{y}(\mathcal{X}))$.

\begin{lemma}\label{lemma2}
Suppose $\delta_{nl}=\delta\geq0$ for all $n\ne l$, $f_n\geq0$, and $\epsilon_{nl}\geq0$. Then $f(\mathcal{X})$ is supermodular over nonempty deployment sets.
\end{lemma}

Under uniform $\delta$, the assignment-dependent cost for client $m$ is $\min_{n\in\mathcal{X}}\zeta_{mn}+\omega\delta(|\mathcal{X}|-1)$. The minimum term is supermodular, the cardinality and deployment terms are modular, and the nonnegative pairwise term is supermodular. Their sum is therefore supermodular.

For large instances, we maximize the submodular surrogate $g(\mathcal{X})=-f(\mathcal{X})$ using the randomized double-greedy heuristic in Algorithm \ref{alg:random}. Because feasible deployments exclude the empty set and $g$ need not be nonnegative, the standard $1/2$ guarantee for unconstrained nonnegative submodular maximization \cite{BuchbinderFNS15} does not directly apply. We therefore use the method as a scalable heuristic and evaluate its solution quality empirically. The algorithm seeds both sets with the best singleton, processes every remaining candidate, and maintains $X_i^s\subseteq Y_i^s$ until the two sets coincide.

\begin{algorithm}[!htbp]
	\DontPrintSemicolon
	\normalem
	\caption{\small{Allocation Approximation Algorithm}} \label{alg:random}
	\footnotesize{
		\KwInput{Candidate set $\mathcal{S}$ and deployment cost $f$}
		\KwOutput{A nonempty deployment set $X^s$}
		$u^*\leftarrow\arg\min_{u\in\mathcal{S}}f(\{u\})$; $X_0^s\leftarrow\{u^*\}$; $Y_0^s\leftarrow\mathcal{S}$\;
		Order $\mathcal{S}\setminus\{u^*\}$ as $u_1,\ldots,u_{z-1}$; set $g\leftarrow-f$\;
		\For{$i=1$ \textbf{to} $z-1$}{\label{A1l1}
			$a_i\leftarrow g(X_{i-1}^s\cup\{u_i\})-g(X_{i-1}^s)$\;
			$b_i\leftarrow g(Y_{i-1}^s\setminus\{u_i\})-g(Y_{i-1}^s)$\;
			$a_i'\leftarrow\max\{a_i,0\}$; $b_i'\leftarrow\max\{b_i,0\}$\;
			$p_i\leftarrow1/2$ if $a_i'+b_i'=0$; otherwise $p_i\leftarrow a_i'/(a_i'+b_i')$\;
			Draw $\rho_i\sim\mathcal{U}[0,1]$\;
			\If{$\rho_i\leq p_i$}{\label{A1l5}
				$X_i^s\leftarrow X_{i-1}^s\cup\{u_i\}$; $Y_i^s\leftarrow Y_{i-1}^s$\;
			}
			\Else{
				$X_i^s\leftarrow X_{i-1}^s$; $Y_i^s\leftarrow Y_{i-1}^s\setminus\{u_i\}$\;
			}
		}
		\textbf{return} $X_{z-1}^s$ \label{A1l9}
	}
\end{algorithm}

\subsection{Rate-Based Routing Protocol Design}\label{protocol}
\textbf{Formal constraints.} Let $\mathcal{D}=\{(s,e):s\ne e,\ d_{se}>0\}$ be the set of active payment demands, and let $\mathbb{P}_{se}$ be the candidate paths for demand $(s,e)$. For path $p$, let $r_p\geq0$ denote its payment rate, and define
\begin{equation}
R_{se}=\sum_{p\in\mathbb{P}_{se}}r_p,\qquad
r_{ab}=\sum_{(s,e)\in\mathcal{D}}\sum_{\substack{p\in\mathbb{P}_{se}\\(a,b)\in p}}r_p.
\label{eq:aggregate-rates}
\end{equation}
For channel $\{a,b\}$, let $b_{ab}$ and $b_{ba}$ be the current directional balances, with $b_{ab}+b_{ba}=c_{ab}$; let $b_{\min}$ be the liquidity reserve and $c_{ab}^{\rm eff}$ the amount available to in-flight TUs. During confirmation delay $\Delta$, SHARE solves
\begin{align}
\max_{\{r_p\}}\quad
&\sum_{(s,e)\in\mathcal{D}}\log R_{se} \label{eq:utility-obj}\\
\text{s.t.}\quad
&R_{se}\Delta\leq d_{se}, &&\forall(s,e)\in\mathcal{D}, \label{demand}\\
&r_{ab}+r_{ba}\leq c_{ab}^{\rm eff}/\Delta, &&\forall\{a,b\}\in\mathbb{E}, \label{capacity}\\
&r_{ab}-r_{ba}\leq(b_{ab}-b_{\min})/\Delta, &&\forall\{a,b\}\in\mathbb{E}, \label{balance-ab}\\
&r_{ba}-r_{ab}\leq(b_{ba}-b_{\min})/\Delta, &&\forall\{a,b\}\in\mathbb{E}, \label{balance-ba}\\
&r_p\geq0, &&\forall p.
\end{align}
The capacity constraint limits total in-flight value, while Equations \eqref{balance-ab} and \eqref{balance-ba} prevent either direction from depleting its reserve.

Each smooth node applies projected primal--dual updates. Let $\lambda_{ab}$ be the capacity multiplier, $\mu_{ab}$ and $\mu_{ba}$ the directional-liquidity multipliers, and $\nu_{se}$ the demand multiplier. With step sizes $\kappa$, $\eta$, and $\chi$, respectively,
\begin{align}
\lambda_{ab}^{+}&=\left[\lambda_{ab}+\kappa\left(r_{ab}+r_{ba}-c_{ab}^{\rm eff}/\Delta\right)\right]_{+}, \label{eq:dual-cap}\\
\mu_{ab}^{+}&=\left[\mu_{ab}+\eta\left(r_{ab}-r_{ba}-(b_{ab}-b_{\min})/\Delta\right)\right]_{+}, \label{eq:dual-ab}\\
\mu_{ba}^{+}&=\left[\mu_{ba}+\eta\left(r_{ba}-r_{ab}-(b_{ba}-b_{\min})/\Delta\right)\right]_{+}, \label{eq:dual-ba}\\
\nu_{se}^{+}&=\left[\nu_{se}+\chi(R_{se}\Delta-d_{se})\right]_{+}. \label{eq:dual-demand}
\end{align}
The signed directional prices are $\xi_{ab}=\lambda_{ab}+\mu_{ab}-\mu_{ba}$ and $\xi_{ba}=\lambda_{ab}+\mu_{ba}-\mu_{ab}$. A negative signed price acts as a rebalancing credit. The charged service fee remains nonnegative: $\textsf{fee}_{ab}=T_{\rm fee}[\xi_{ab}]_{+}$. We define the effective link price as $\widehat{\xi}_{ab}=\xi_{ab}+\textsf{fee}_{ab}$ and the path price as $\varrho_p=\sum_{(a,b)\in p}\widehat{\xi}_{ab}$. For logarithmic utility, the projected path-rate update is
\begin{equation}
r_p^{+}=\left[r_p+\alpha\left(\frac{1}{R_{se}}-\nu_{se}-\varrho_p\right)\right]_{+},
\label{eq:rate-update}
\end{equation}
for $p\in\mathbb{P}_{se}$. Global state from the preceding epoch initializes these prices, while current requests determine the local updates. Each payment is split over at most $K$ selected paths, consistent with the $K$ TUs defined in Section \ref{sec_Problem}.

\textbf{Congestion control.} Price updates react to persistent load but may not immediately absorb queue bursts. SHARE therefore maintains a bounded window $w_p\in[w_{\min},w_{\max}]$ for each path. If queue delay exceeds $T$ or queued value exceeds the window, the smooth node decreases the window by $\beta$ and caps the sending rate. After successful transmission, it increases the window additively using $\gamma$. Algorithm \ref{alg:RP} summarizes one update epoch.

Compared with Spider \cite{Sivaraman2020HighTC}, SHARE differs in three respects. It incorporates PCH service fees, combines rate control with queue-based congestion control, and delegates route computation to attested PCH enclaves rather than requiring senders to compute complete paths.

\begin{algorithm}[!htbp]
	\DontPrintSemicolon
	\normalem
	\caption{\small{Distributed Routing Decision Protocol}} \label{alg:RP}
	\footnotesize{
		\KwInput{Demands $\{d_{se}\}$, paths $\{\mathbb{P}_{se}\}$, balances, rates, prices, and windows}
		\KwOutput{Updated path rates $\{r_p\}$}
		Compute $R_{se}$ and aggregate channel rates $r_{ab}$ using Equation \eqref{eq:aggregate-rates}\;
		\ForEach{$(s,e)\in\mathcal{D}$}{
			Update $\nu_{se}$ using Equation \eqref{eq:dual-demand}\;
		}
		\ForEach{$\{a,b\}\in\mathbb{E}$}{
			Update $\lambda_{ab}$, $\mu_{ab}$, and $\mu_{ba}$ using Equations \eqref{eq:dual-cap}--\eqref{eq:dual-ba}\;
			$\xi_{ab}\leftarrow\lambda_{ab}+\mu_{ab}-\mu_{ba}$; $\xi_{ba}\leftarrow\lambda_{ab}+\mu_{ba}-\mu_{ab}$\;
			For $uv\in\{ab,ba\}$, set $\textsf{fee}_{uv}\leftarrow T_{\rm fee}[\xi_{uv}]_{+}$ and $\widehat{\xi}_{uv}\leftarrow\xi_{uv}+\textsf{fee}_{uv}$\;
		}
		\ForEach{$(s,e)\in\mathcal{D}$}{
			Select at most $K$ paths from $\mathbb{P}_{se}$ and split the payment into TUs\;
			\ForEach{selected path $p\in\mathbb{P}_{se}$}{
				$\varrho_p\leftarrow\sum_{(a,b)\in p}\widehat{\xi}_{ab}$; update $r_p$ using Equation \eqref{eq:rate-update}\;
				\If{$q_p^{\rm delay}>T$ \textbf{or} $q_p^{\rm amount}>w_p$}{
					$w_p\leftarrow\max\{w_{\min},w_p-\beta\}$;
					$r_p\leftarrow\min\{r_p,w_p/\Delta\}$\;
				}
				\ElseIf{the TU is transmitted}{
					$w_p\leftarrow\min\{w_{\max},w_p+\gamma/\sum_{p'\in\mathbb{P}_{se}}w_{p'}\}$\;
				}
			}
		}
		\textbf{return} all updated rates $\{r_p\}$
	}
\end{algorithm}

\subsection{Channel Concurrency}\label{Concurrency}
\textbf{Concurrent channels.} Multi-hop payments temporarily lock liquidity on intermediate channels, which can limit concurrent forwarding \cite{LN, RD}. SHARE does not bypass the blockchain funding requirement for a base channel. Instead, smooth nodes pre-fund inter-PCH capacity on-chain and use attested off-chain state transitions to partition that capacity into concurrently processed logical channels or subchannels, analogous to Teechain's off-chain execution model \cite{LindNEKSP19}. This design reduces online coordination latency without creating unfunded payment capacity.

In the Lightning protocol, peers negotiate the \texttt{max\_accepted\_htlcs} channel parameter, whose specified upper bound is 483\footnote{https://github.com/lightning/bolts/blob/master/02-peer-protocol.md}. Thus, increasing only the monetary capacity of one channel does not increase its number of simultaneous HTLC slots. Multiple pre-funded channels or logical processing lanes can expose more parallelism, but excessive concurrency also increases contention. We characterize this tradeoff through channel concurrency benefit theory (CCBT).

The number of channels that a smooth node can process concurrently depends on both workload and available resources. Workload consists of client payment requests. Resources include SGX processor and enclave memory, pre-funded channel liquidity, and TU queue capacity. Concurrent requests contend for these resources, while smooth nodes also synchronize shared state. We model these effects using the Universal Scalability Law (USL) \cite{USL}.

Let $N_{\rm CC}$ denote the number of concurrent channels. The system throughput is
\begin{equation}
T(N_{\rm CC})=
\frac{\varepsilon N_{\rm CC}}
{1+\varsigma(N_{\rm CC}-1)+\varpi N_{\rm CC}(N_{\rm CC}-1)},
\label{CCBT}
\end{equation}
where $\varepsilon>0$ is the single-channel throughput $T(1)$, $\varsigma\geq0$ captures contention for serialized resources, and $\varpi\geq0$ captures coherence and synchronization overhead. Equation \eqref{CCBT} remains positive, but its marginal gain can become negative as $N_{\rm CC}$ grows. When $\varpi>0$ and $\varsigma<1$, the continuous relaxation is maximized at $N_{\rm CC}^{*}=\sqrt{(1-\varsigma)/\varpi}$; an implementation selects a nearby feasible integer and adapts it to the observed workload.

\textbf{Batch processing.} SHARE batches state updates and seals the resulting checkpoints to reduce enclave bookkeeping and rollback-protection overhead \cite{matetic2017rote}. Legacy SGX platform services exposed rate-limited hardware monotonic counters; therefore, the design does not increment a hardware counter for every TU. Instead, it follows Teechain \cite{LindNEKSP19} by aggregating updates for each sender--receiver pair and commits one authenticated checkpoint per batch. The security argument requires freshness and rollback detection, not a specific counter throughput.

\section{Security Analysis} \label{Security}
This section gives a property-based security analysis using a UC-inspired ideal functionality \cite{Canetti2001}. We specify the real protocol $\textbf{Prot}_{SHARE}$ and the leakage and external behavior captured by $\mathcal{F}_{SHARE}$. The theorem establishes three scoped properties rather than a general UC realization; Appendix D provides the complete argument.

The static PPT adversary may observe the network, control corrupted smooth-node hosts, and drop, delay, replay, or reorder their messages. It corrupts at most $t$ of $\iota\geq2t+1$ KMG members before setup, where $\tau_{\rm dec}=t+1$, but cannot break honest TEEs, extract honest shares, bypass enclave-bound decryption authorization, or forge attestation evidence. The analysis covers routing confidentiality, execution integrity, and receiver-side acceptance atomicity under explicit leakage; it excludes simultaneous path resolution under arbitrary scheduling. Relationship unlinkability remains conditional on the underlying PCH construction.

The ideal functionality $\mathcal{F}_{SHARE}$ specifies protected inputs, permitted leakage, complete-set acceptance, and delayed outputs.

\subsection{The Real-World Protocol Model \texorpdfstring{$\textbf{Prot}_{SHARE}$}{Prot-SHARE}}
$\textbf{Prot}_{SHARE}$ operates in the $(\mathcal{G}_{\rm{att}}, \mathcal{F}_{\rm{bc}})$-hybrid model. A $t$-private, robust threshold scheme $\mathcal{TE}=(\textsf{DKG},\textsf{Enc},\textsf{PDec},\textsf{VerifyShare},\textsf{Combine})$ supplies one-time key entries. KMG members release verifiable shares only to an authorized attested program, which combines at least $\tau_{\rm dec}$ shares without reconstructing a complete private key. A shared hashlock and the Basic MPP terminal rule restrict an honest receiver to complete-set acceptance.

Appendix D gives the protocol algorithms and states the scoped security theorem.

\subsection{The Ideal-World Functionality \texorpdfstring{\textcolor{brickred}{\ensuremath{\mathcal{F}_{SHARE}}}}{F-SHARE}}
\textcolor{brickred}{\ensuremath{\mathcal{F}_{SHARE}}} separates confidential payloads from the identifiers, lengths, timing, routing metadata, and HTLC outcomes permitted by the leakage model.

Each participant may act as a client or smooth node over secure channels. For a protected message $m$, the adversary learns only $\ell(m)$. Under the standard delayed-output convention \cite{Canetti2001}, outputs pass through the ideal adversarial interface before delivery.

Initialization exposes only its declared leakage. Processing reveals $\ell(D_{\textcolor{colorwheel}{\textsf{tid}}})$ and the TU leakages, and records receiver acceptance only after every TU is locked and a valid common preimage is supplied. An incomplete set is not accepted; subsequent per-TU fulfillment or timeout remains visible through the HTLC interface. The sender receives a proof and acknowledgment only after all TU fulfillments are observed.

Under the stated cryptographic, attestation, and HTLC assumptions, Theorem 1 establishes routing confidentiality, execution integrity, and receiver-side acceptance atomicity against static corruption of at most $t$ KMG members.

\section{Evaluation}\label{Section:Evaluation}
\begin{figure*}[!tbp]
	\centering
	\subfloat[Influence of the channel size]{
		\begin{minipage}[t]{0.23\linewidth}
			\centering
			\includegraphics[width=1.4in]{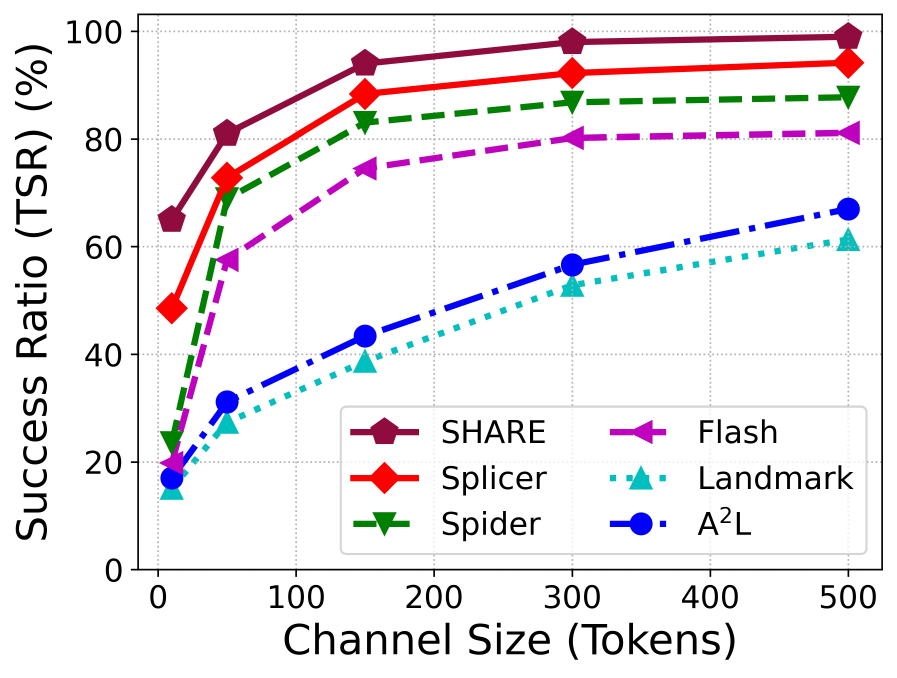}
			\label{fig_channelSize-successRatio}
		\end{minipage}%
	}%
	\hspace{1mm}
	\subfloat[Influence of the transaction size]{
		\begin{minipage}[t]{0.23\linewidth}
			\centering
			\includegraphics[width=1.4in]{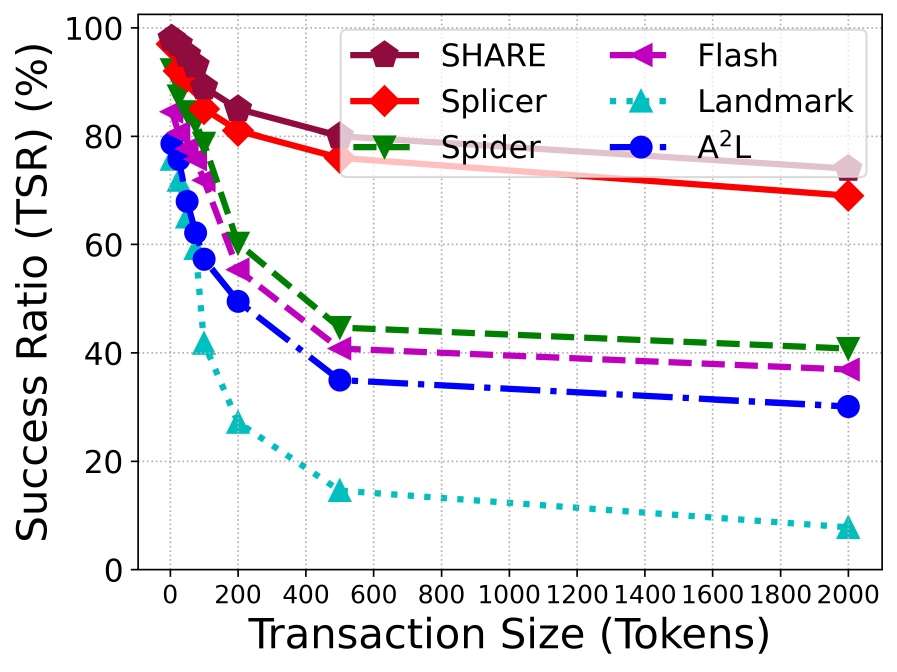}
			\label{fig_TransactionSize-successRatio}
		\end{minipage}%
	}%
	\hspace{1mm}
	\subfloat[Influence of the update time]{
		\begin{minipage}[t]{0.23\linewidth}
			\centering
			\includegraphics[width=1.4in]{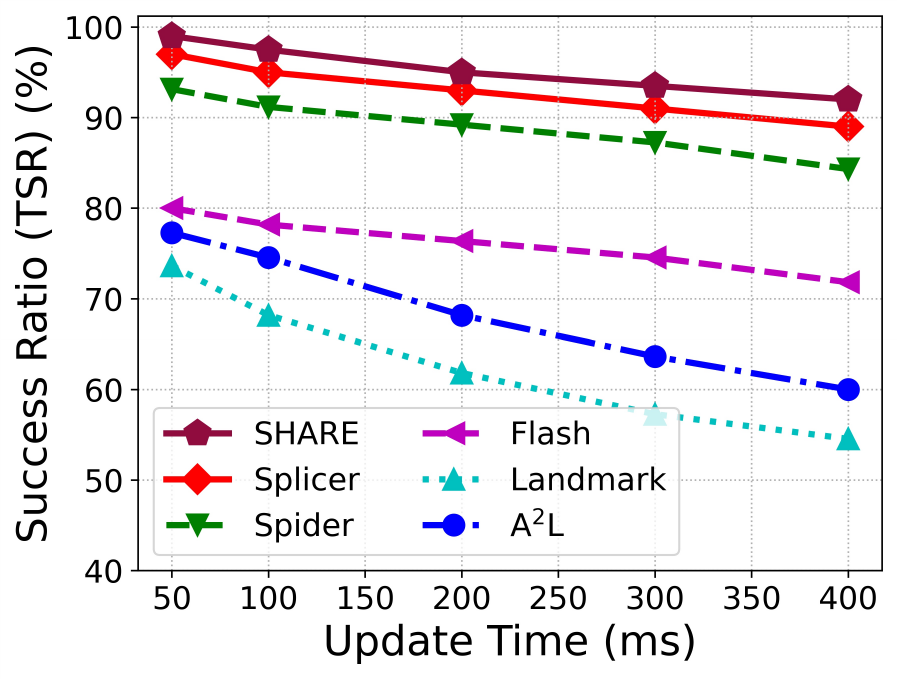}
			\label{fig_updateTime-successRatio}
		\end{minipage}
	}%
	\hspace{1mm}
	\subfloat[Normalized throughput]{
		\begin{minipage}[t]{0.23\linewidth}
			\centering
			\includegraphics[width=1.4in]{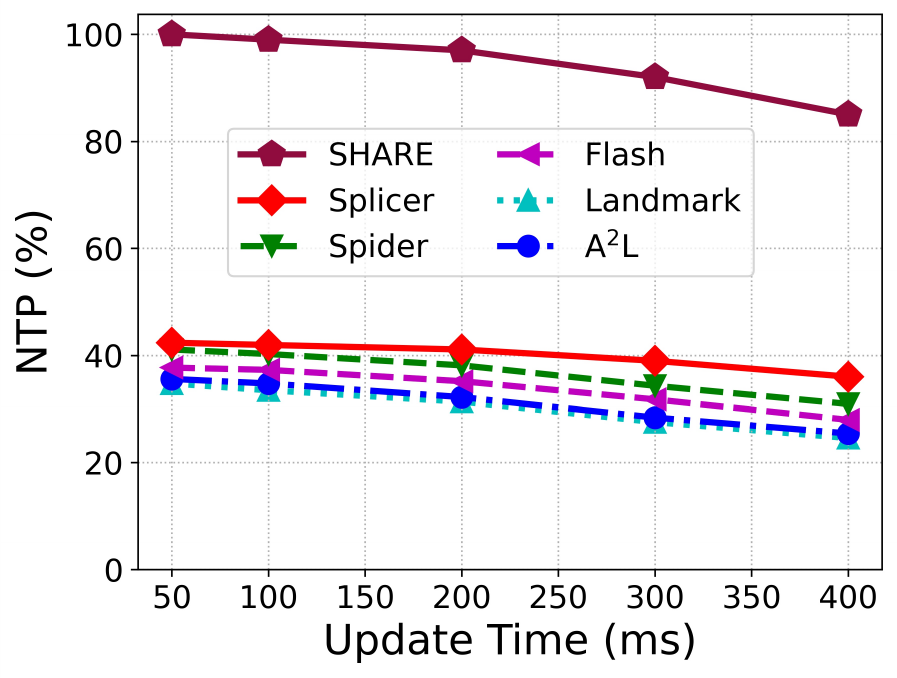}
			\label{fig_updateTime-throughput}
		\end{minipage}
	}%
	\centering
	\caption{Simulation comparison with non-TEE schemes in small-scale networks. Curves report descriptive point estimates.}
	\label{performance}
\end{figure*}

\begin{figure*}[!tbp]
	\centering
	\subfloat[Influence of the channel size]{
		\begin{minipage}[t]{0.23\linewidth}
			\centering
			\includegraphics[width=1.4in]{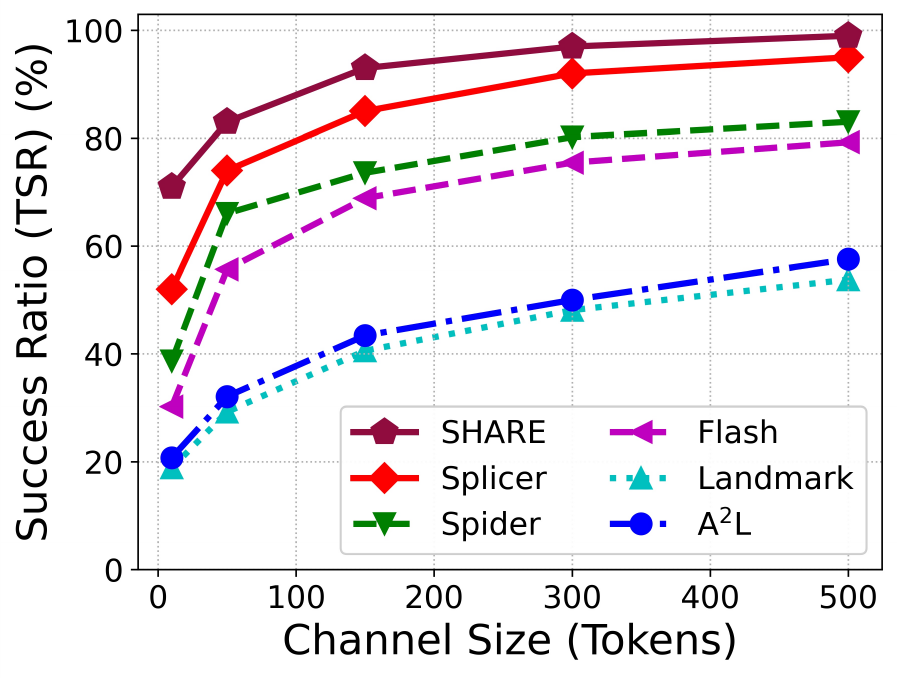}
			\label{fig_channelSize-successRatio2}
		\end{minipage}%
	}%
	\hspace{1mm}
	\subfloat[Influence of the transaction size]{
		\begin{minipage}[t]{0.23\linewidth}
			\centering
			\includegraphics[width=1.4in]{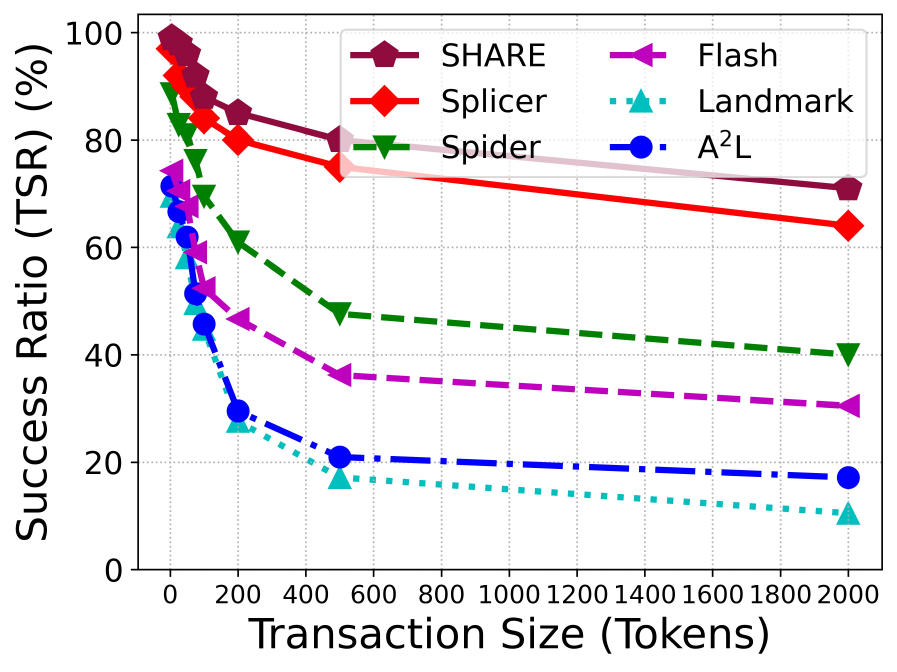}
			\label{fig_TransactionSize-successRatio2}
		\end{minipage}%
	}%
	\hspace{1mm}
	\subfloat[Influence of the update time]{
		\begin{minipage}[t]{0.23\linewidth}
			\centering
			\includegraphics[width=1.4in]{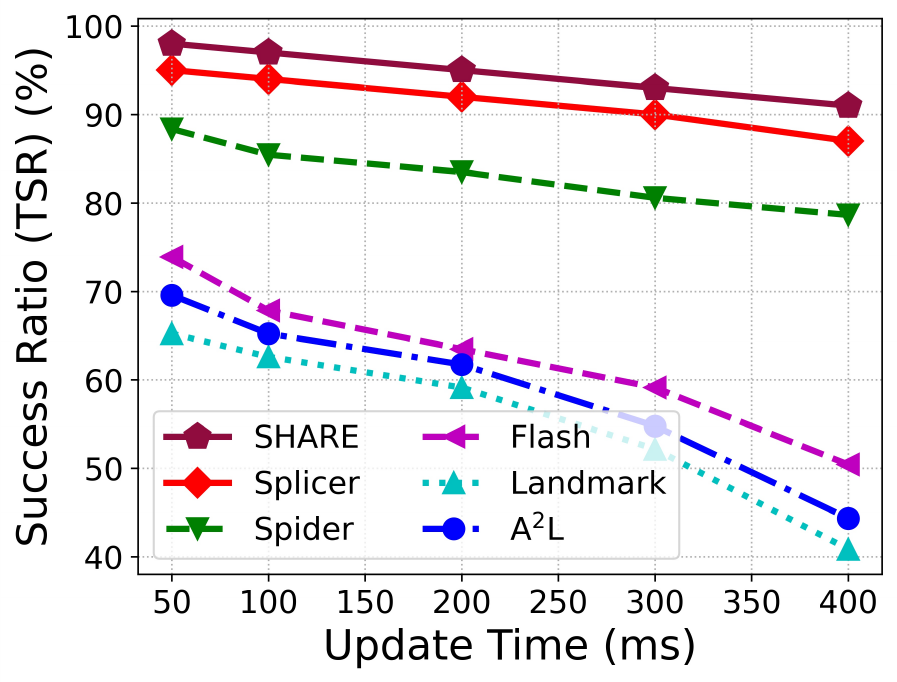}
			\label{fig_updateTime-successRatio2}
		\end{minipage}
	}%
	\hspace{1mm}
	\subfloat[Normalized throughput]{
		\begin{minipage}[t]{0.23\linewidth}
			\centering
			\includegraphics[width=1.4in]{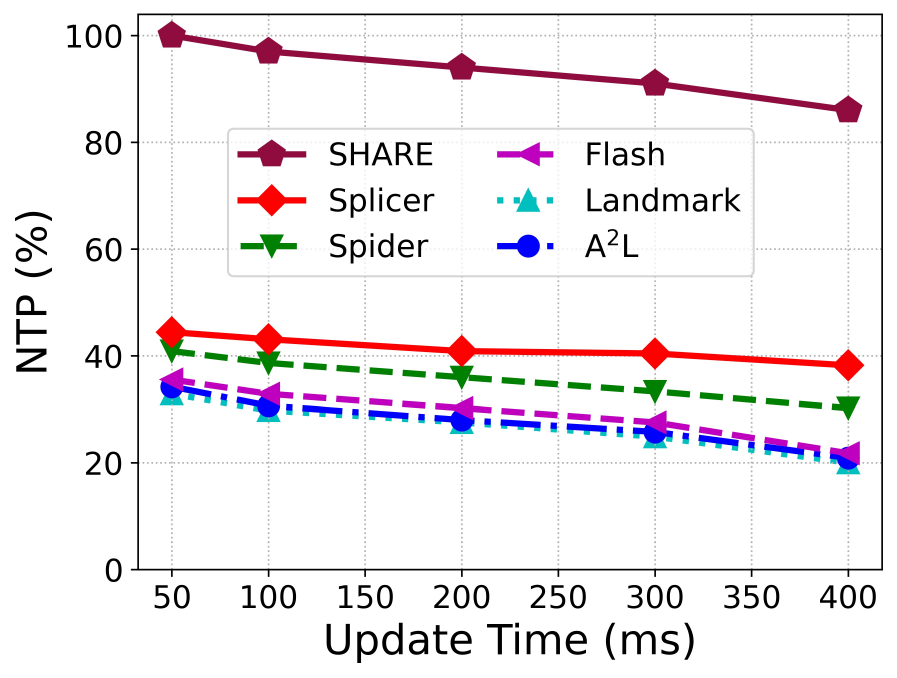}
			\label{fig_updateTime-throughput2}
		\end{minipage}
	}%
	\centering
	\caption{Simulation comparison with non-TEE schemes in large-scale networks. Curves report descriptive point estimates.}
	\label{performance2}
\end{figure*}

\subsection{Experiment Setup}
Our evaluation uses MATLAB simulations and an LND/SGX prototype. Fig. \ref{performance}, Fig. \ref{performance2}, Fig. \ref{allocation}, and Table \ref{tab-choices} are simulation results; Fig. \ref{Fig-TEE} and Fig. \ref{channel-tps} are produced by a prototype that executes routing and concurrent-channel logic inside SGX. We simulate networks with 100 and 3,000 nodes. The prototype runs on an Intel i7-9750H CPU with SGX support, 32 GB RAM, a 500 GB SSD, and a 10 Gbps network card. Following Spider's benchmark, we generate Watts--Strogatz topologies using ROLL \cite{HadianNMQ16}; capacities follow the heavy-tailed Lightning distribution \cite{TikhomirovMM20}, transaction directionality is derived from a preprocessed Lightning dataset, and amounts follow the credit-card dataset \cite{Credit}.

\textbf{Parameter settings.} The minimum, average, and median channel capacities are 10, 403, and 152 tokens. The timeout is 3 seconds, TU values range from 1 to 4 tokens, and $K=5$. We use $\zeta_{mn}=0.02\,hops_{mn}$, $\delta_{nl}=0.01\,hops_{nl}$, and $\epsilon_{nl}=0.05\,hops_{nl}$. The allocation experiment sets $f_n=0$, focusing on the restricted management--synchronization objective. Each queue holds 8,000 tokens, $\beta=10$, $\gamma=0.1$, $\Delta_{\rm upd}=200$ ms, $T=400$ ms, $N_{\rm CC}=5$, and the default batch wait is 100 ms.

\textbf{Evaluation scope.} The prototype evaluates TEE-assisted routing and channel concurrency, but excludes a production threshold-DKG service and KMG partial-decryption communication. Reported latency and throughput therefore exclude offline pool-refill cost and online $\mathcal{TE}.\textsf{PDec}$ network overhead. The retained artifacts contain aggregate curves, so the following results are descriptive point estimates without confidence intervals.

Let $\mathcal{T}$ be the generated transactions and $\mathcal{T}_{\rm succ}$ the completed subset. We use
\begin{equation}
\mathrm{TSR}=\frac{|\mathcal{T}_{\rm succ}|}{|\mathcal{T}|}\times100\%,\qquad
\mathrm{NTP}=\frac{\sum_{t\in\mathcal{T}_{\rm succ}}\mathrm{val}(t)}{\sum_{t\in\mathcal{T}}\mathrm{val}(t)}\times100\%.
\label{eq:evaluation-metrics}
\end{equation}
TSR measures completion frequency, whereas normalized throughput (NTP) weights completion by payment value. Average gains over operating points $\Omega$ are computed as $|\Omega|^{-1}\sum_{o\in\Omega}(M_A(o)-M_B(o))/M_B(o)\times100\%$.

\subsection{Performance of SHARE} \label{Per-Spl}
We evaluate SHARE along three axes: performance relative to non-TEE schemes, throughput and latency relative to TEE-assisted baselines, and the effect of concurrent channels. The comparison focuses on the mechanisms that SHARE changes: multi-PCH coordination, distributed routing, and TEE-assisted execution.

\subsubsection{Comparison with Non-TEE Schemes}
The MATLAB comparison uses the same simulator, topology, capacity assignment, trace, timeout, and TU constraints for all non-TEE routing rules. \textit{Splicer} \cite{yang2023optimal} is SHARE's closest non-TEE ablation, while \textit{Spider} \cite{Sivaraman2020HighTC}, \textit{Flash} \cite{WangXJW19}, \textit{Flare} \cite{Flare2016}, and \textit{A$^2$L} \cite{Tairi} represent source-routing, landmark, and single-hub designs.

As shown in Fig. \ref{performance} and Fig. \ref{performance2}, SHARE achieves the highest TSR at both network scales in the tested settings. Its average relative gain over the comparison set is 51.1\% when channel size varies and 43.6\% when transaction size varies. As the update interval increases, all methods complete fewer transactions before timeout, but SHARE remains above 91.0\% TSR and exceeds Spider by 7.2\% and 13.8\% at the two scales. It also exceeds A$^2$L by 38.8\% and 60.3\%, with a 78.2\% average advantage over A$^2$L. Relative to Splicer, SHARE improves TSR by 6.3\% on average.

Using Equation \eqref{eq:evaluation-metrics}, SHARE achieves an average NTP gain of 181.5\% over the comparison set; the corresponding large-scale gain is 189.3\%. Relative gains over Spider are 156.0\% and 161.3\% at the two scales, and gains over A$^2$L are 202.5\% and 235.4\%. SHARE also exceeds Splicer by 130.9\% on average, showing that it completes a larger fraction of payment value under the tested contention and timeout settings.

Overall, the simulations support the claim that collaborative PCH routing improves completion and payment-value throughput in the evaluated workloads.

\begin{figure}[!htbp]
	\centering
	\subfloat[Normalized throughput]{
		\begin{minipage}[t]{0.49\linewidth}
			\centering
			\includegraphics[width=1.6in]{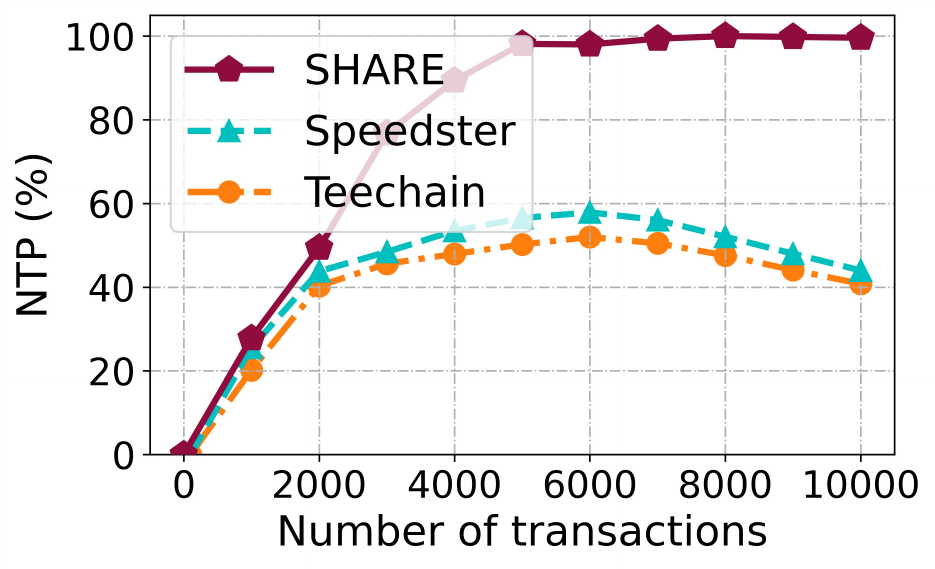}
			\label{TPS-3Type}
		\end{minipage}%
	}%
	\subfloat[Transaction latency]{
		\begin{minipage}[t]{0.49\linewidth}
			\centering
			\includegraphics[width=1.6in]{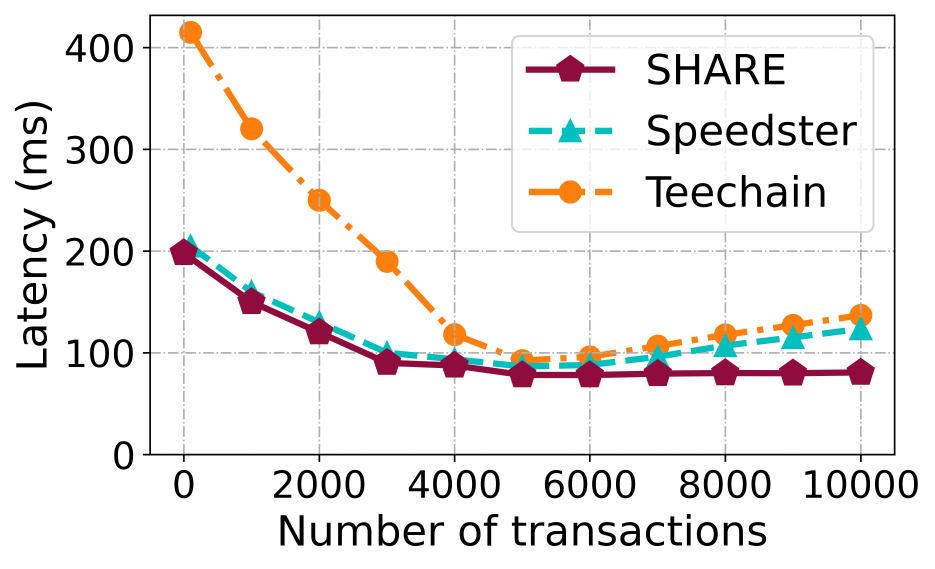}
			\label{Latency-3Type}
		\end{minipage}
	}%
	\centering
	\caption{LND/SGX prototype comparison under a shared transaction trace. Baseline adaptations expose the workload-matched execution path used here.}
	\label{Fig-TEE}
\end{figure}

\begin{figure}[!htbp]
	\centering
	\includegraphics[width=3in]{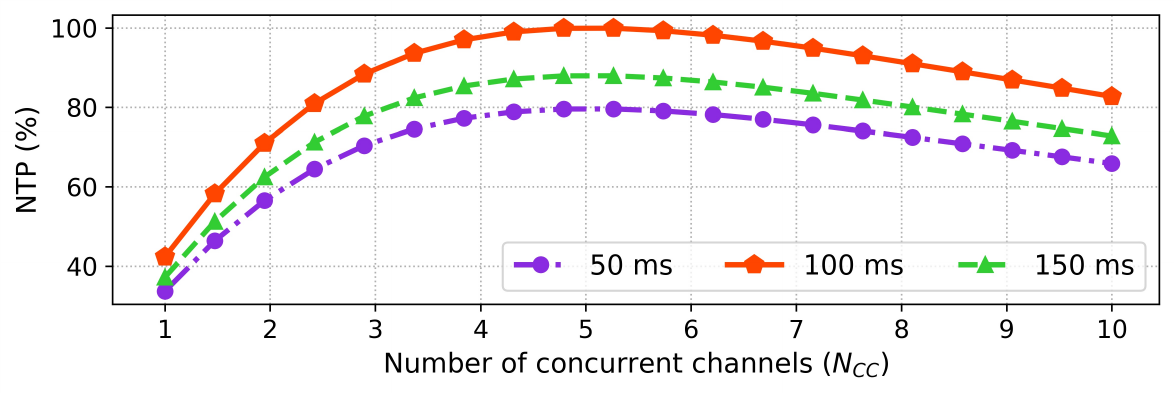}
	\caption{LND/SGX prototype evaluation of concurrent logical channels under different batch waits.}
	\label{channel-tps}
\end{figure}

\begin{figure*}[!tbp]
	\hspace{-7mm}
	\centering
	\subfloat[Balance cost]{
		\begin{minipage}[t]{0.145\linewidth}
			\centering
			\includegraphics[width=1.15in]{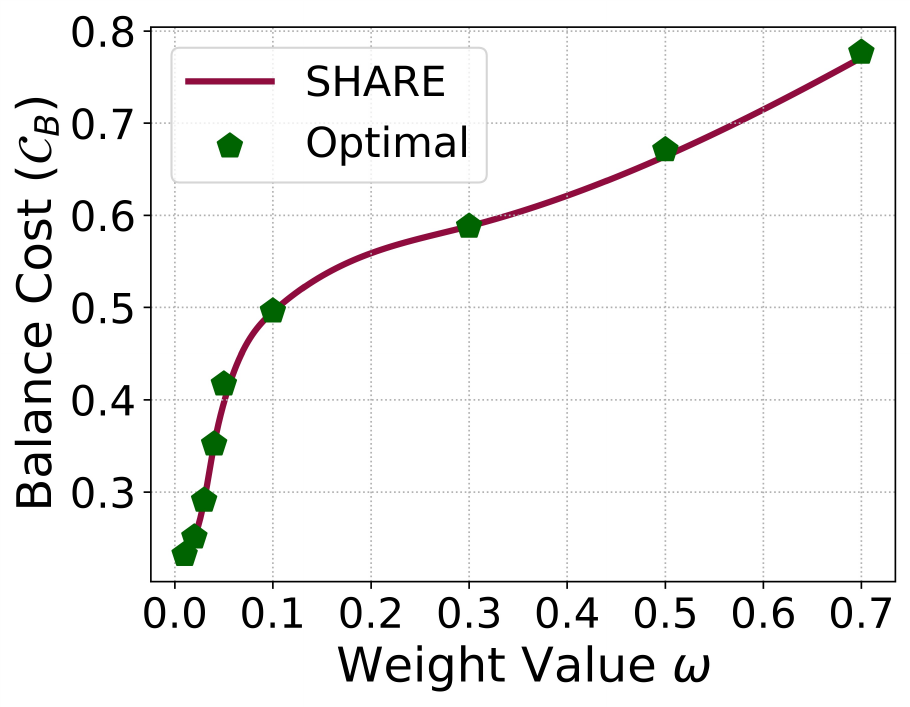}
			\label{fig_para-cost}
		\end{minipage}%
	}%
	\hspace{1mm}
	\subfloat[Tradeoff in costs]{
		\begin{minipage}[t]{0.145\linewidth}
			\centering
			\includegraphics[width=1.15in]{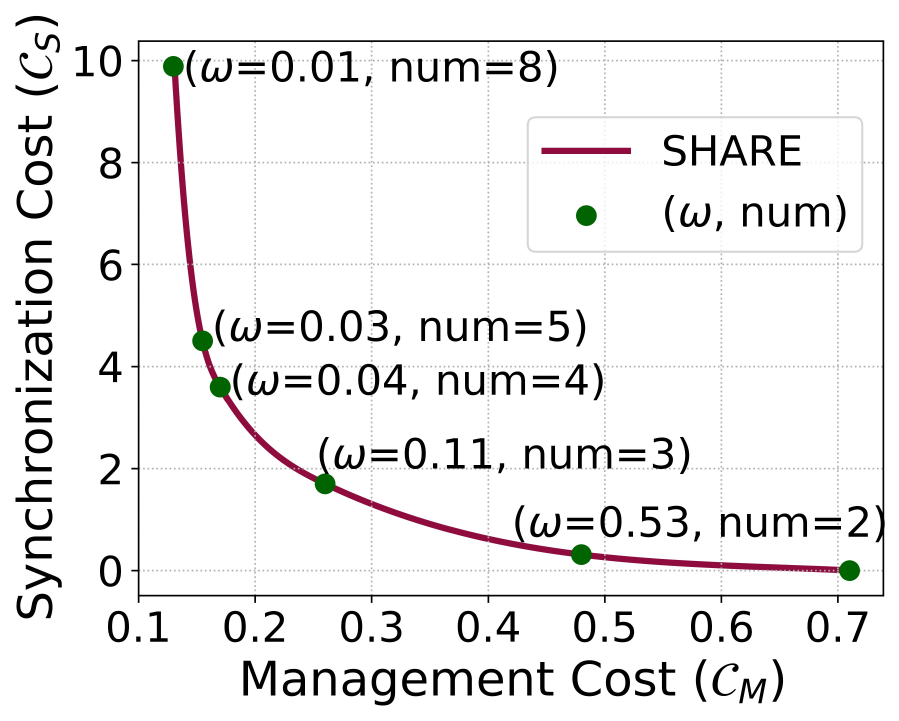}
			\label{fig_tradeoff}
		\end{minipage}%
	}%
	\hspace{1mm}
	\subfloat[Small-scale PCNs]{
		\begin{minipage}[t]{0.145\linewidth}
			\centering
			\includegraphics[width=1.165in]{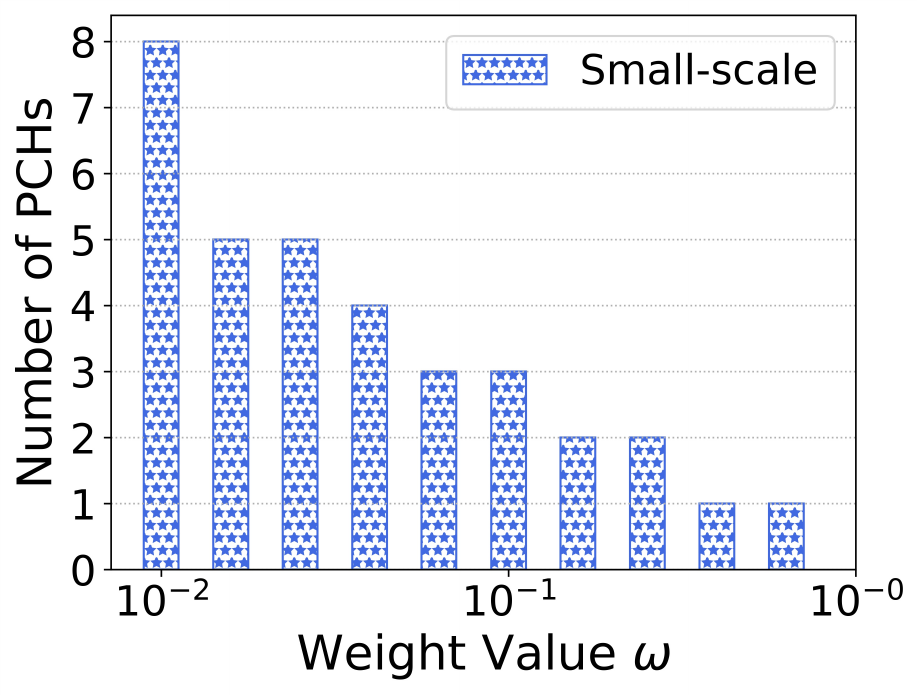}
			\label{number-small}
		\end{minipage}
	}%
	\hspace{1mm}
	\subfloat[Large-scale PCNs]{
		\begin{minipage}[t]{0.145\linewidth}
			\centering
			\includegraphics[width=1.2in]{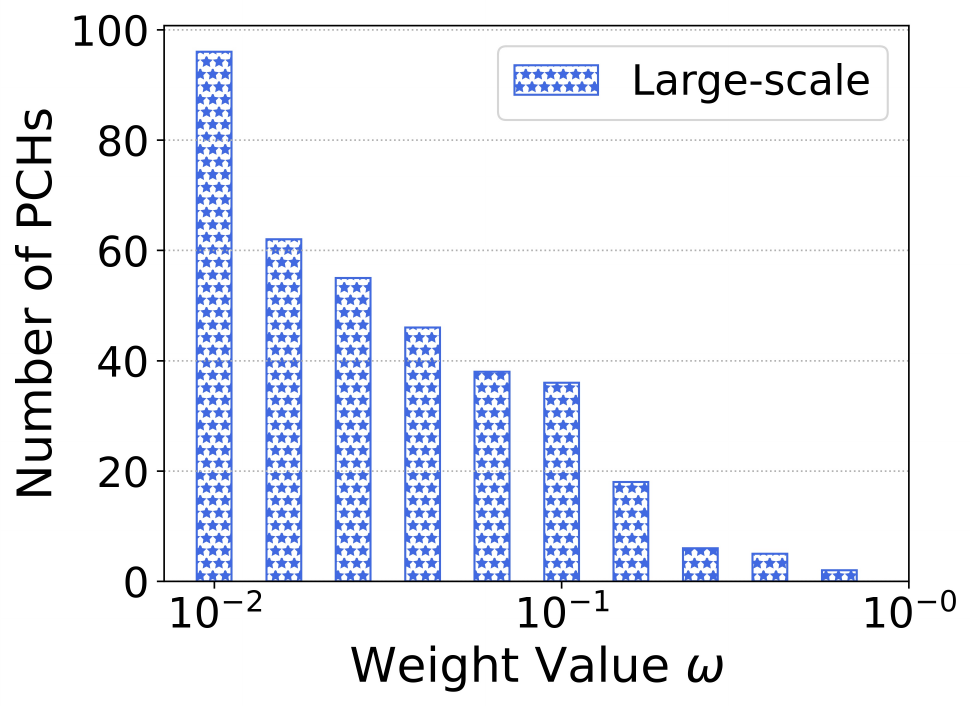}
			\label{number-large}
		\end{minipage}
	}%
	\hspace{1mm}
	\subfloat[Costs in small PCNs]{
		\begin{minipage}[t]{0.145\linewidth}
			\centering
			\includegraphics[width=1.2in]{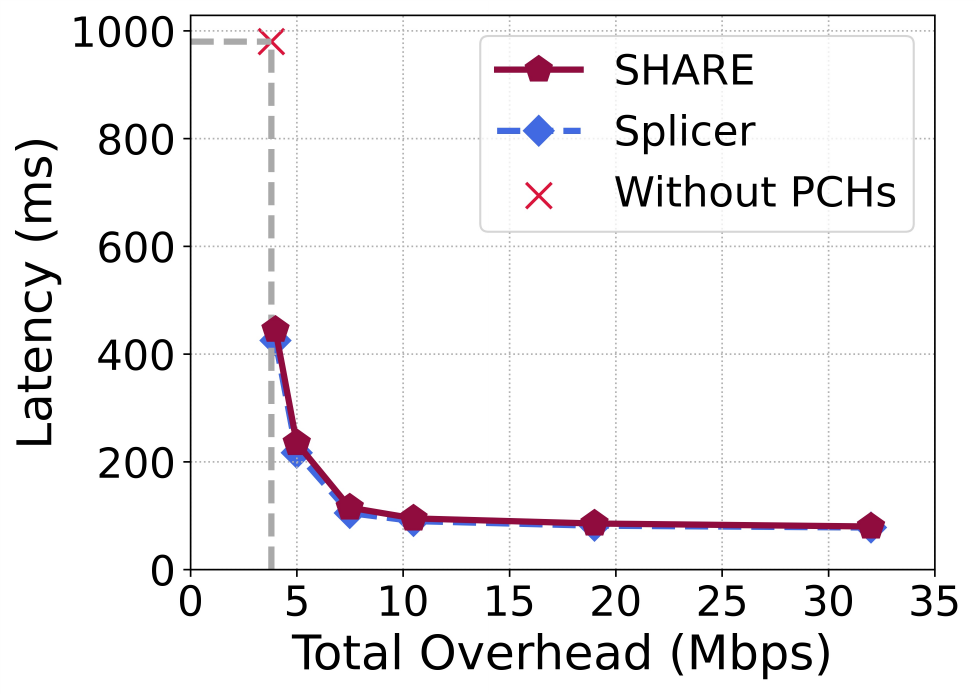}
			\label{cost-small}
		\end{minipage}
	}%
	\hspace{1mm}
	\subfloat[Costs in large PCNs]{
		\begin{minipage}[t]{0.145\linewidth}
			\centering
			\includegraphics[width=1.2in]{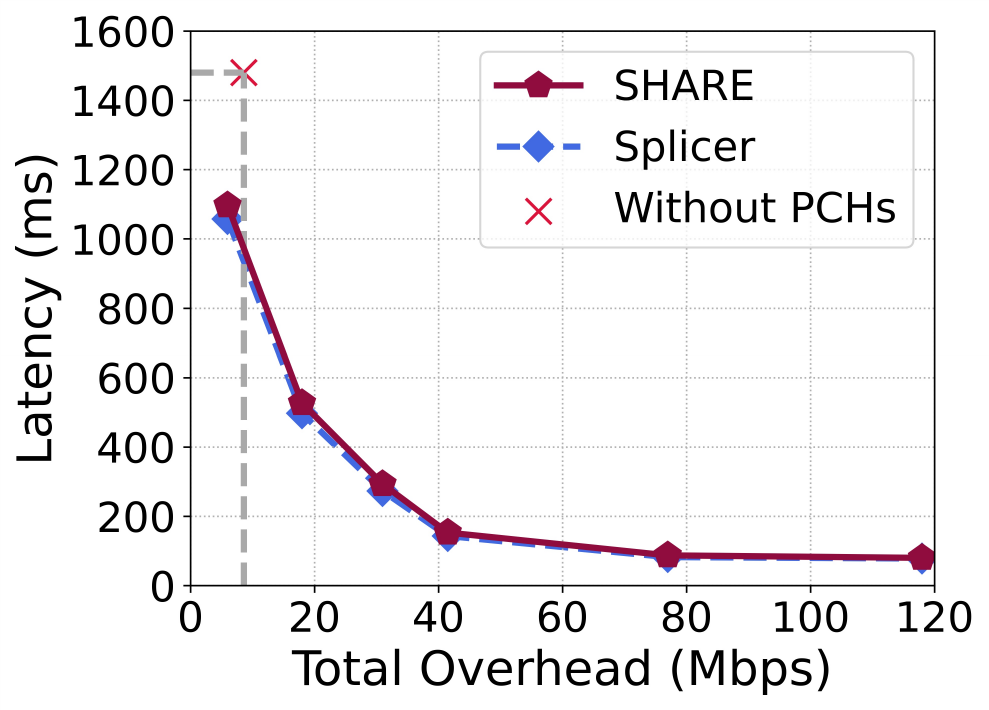}
			\label{cost-large}
		\end{minipage}
	}%
	\centering
	\caption{Simulation of smooth-node allocation under the restricted objective with $f_n=0$.}
	\label{allocation}
\end{figure*}

\begin{table*}[!tbp]
	\centering
	\caption{Routing-strategy sensitivity. Entries report TSR (\%).}
	\label{tab-choices}
	\resizebox{1.9\columnwidth}{!}{
		\begin{tabular}{l cccc cccc cccc}
			\toprule
			\multirow{2}{*}{Scale} & \multicolumn{4}{c}{Path type} & \multicolumn{4}{c}{Number of paths} & \multicolumn{4}{c}{Scheduling policy} \\
			\cmidrule(lr){2-5}\cmidrule(lr){6-9}\cmidrule(lr){10-13}
			& KSP & Heuristic & EDW & EDS & 1 & 3 & 5 & 7 & FIFO & LIFO & SPF & EDF \\
			\midrule
			Small & 66.41 & 77.83 & 86.37 & 82.14 & 34.54 & 72.38 & \textbf{88.54} & 83.74 & 53.81 & \textbf{90.35} & 76.18 & 70.44 \\
			Large & 59.53 & 76.32 & \textbf{91.63} & 85.16 & 36.76 & 67.29 & \textbf{92.29} & 86.30 & 63.48 & \textbf{95.40} & 83.19 & 80.48 \\
			\bottomrule
		\end{tabular}%
	}
\end{table*}%

\subsubsection{Comparison with TEE-Assisted Schemes}
We evaluate NTP and latency under a trace containing up to 10,000 transactions. Because Teechain \cite{LindNEKSP19} and Speedster \cite{Liao2021} have different native system models, Fig. \ref{Fig-TEE} compares workload-matched TEE-protected execution paths rather than complete protocol deployments.

In this prototype, SHARE's average NTP is 72.6\% and 90.9\% higher than the adapted Speedster and Teechain execution paths, respectively. All three curves initially increase with transaction volume, but beyond approximately 2,000 transactions the adapted baselines grow more slowly, whereas SHARE continues to increase over a wider range.

Latency initially decreases as larger batches amortize fixed execution costs. Beyond approximately 5,000 transactions, latency increases for the adapted Speedster and Teechain paths, while SHARE remains more stable, which is consistent with moving route computation away from clients.

These results concern the measured execution paths and exclude the production DKG/PDec service.

\subsubsection{Performance Benefits of Concurrent Channels}
Fig. \ref{channel-tps} shows normalized peak throughput as $N_{\rm CC}$ varies under three batch waits. Throughput first increases and then decreases, reflecting the transition from parallelism gains to contention overhead. A 100 ms wait and $N_{\rm CC}=5$ provide the best measured tradeoff and are used as defaults.

\subsection{Evaluation of Smooth Node Allocation} \label{Eva_SN_Pla}
This subsection evaluates the restricted allocation objective from Section \ref{detail_problem} with $f_n=0$, focusing on the management--synchronization tradeoff and routing behavior under smooth-node deployment.

\subsubsection{Efficiency Tradeoff}
Fig. \ref{fig_para-cost} and Fig. \ref{fig_tradeoff} show how the restricted balance cost changes with $\omega$. The heuristic tracks the exact small-instance solution over the displayed values, and the cost tradeoff follows the expected trend: higher $\omega$ favors fewer smooth nodes and lower synchronization overhead, whereas lower $\omega$ favors more smooth nodes and lower client-management overhead.

Fig. \ref{number-small} and Fig. \ref{number-large} show that the selected number of smooth nodes increases with network scale, matching the need for more coordination capacity as the client population grows.

\subsubsection{Allocation Effectiveness Verification}
Fig. \ref{cost-small} and Fig. \ref{cost-large} compare latency and traffic overhead with and without smooth nodes. In the small-scale setting, SHARE reduces latency by 72.0\% relative to the PCH-free baseline and has 5.5\% higher average latency than Splicer, reflecting the measured overhead of SHARE's simulated TEE-assisted path.

\subsection{Routing Strategy Selection in SHARE} \label{Choices}
We now select the routing primitives used by SHARE for the final configuration. The goal of this subsection is not to show that every candidate is equally good, but to justify the choices that appear in the protocol and the implementation.

\subsubsection{Path Types}
Table \ref{tab-choices} compares $k$-shortest paths (KSP), the heuristic selector, edge-disjoint widest paths (EDW), and edge-disjoint shortest paths (EDS). EDW performs best at both scales, indicating that wide disjoint paths use heavy-tailed channel liquidity more effectively than hop-count optimization alone.

\subsubsection{Number of Paths}
TSR increases when moving from 1 to 3 to 5 EDW paths, but saturates and slightly drops at 7 paths. We therefore set the default path count to 5, balancing load spreading and routing overhead.

\subsubsection{Scheduling Algorithms}
We compare FIFO, LIFO, shortest-payment-first (SPF), and earliest-deadline-first (EDF) scheduling. LIFO yields the highest TSR at both scales; SPF may defer large payments, while EDF prioritizes deadlines without directly considering liquidity. SHARE therefore uses LIFO in the evaluated configuration.

In practice, these routing choices can be adjusted for target workloads.

\section{Conclusion} \label{conclu}
This paper shows that PCHs can be organized as distributed service infrastructure rather than isolated payment relays. SHARE coordinates multiple TEE-enabled smooth nodes to jointly allocate service load, route concurrent payments, and protect sensitive routing information. The main finding is that multi-PCH coordination improves both scalability and service robustness: load-aware allocation reduces concentration at individual hubs, rate-controlled multipath routing preserves channel liquidity under contention, and TEE-assisted execution supports confidential coordination without requiring every client to run trusted hardware. The security analysis establishes routing confidentiality, execution integrity, and receiver-side acceptance atomicity under the stated leakage and threshold-corruption assumptions.

The evaluation further indicates that this architecture can translate the above design goals into measurable service benefits, including higher transaction completion and larger completed payment value under the tested workloads. These results suggest that secure multi-hub coordination is a viable direction for building scalable, privacy-preserving off-chain payment services. Future work will extend SHARE with production-grade asynchronous threshold-key management and broader deployment studies on longitudinal Lightning snapshots, heterogeneous hardware, and more dynamic workload and failure settings.

\appendices
\section{Background and Preliminaries}\label{pre}
This appendix provides background on payment-channel liquidity and trusted execution.

\subsection{Payment Channel Network}
\begin{figure}[h]
	\vspace{-0.2cm}
	\subfloat[A simple PCN.]{
		\begin{minipage}[t]{0.33\linewidth}
			\centering
			\includegraphics[width=1in]{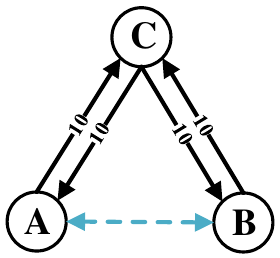}
			\label{fig_PCN1}
		\end{minipage}%
	}%
	\subfloat[The initial state.]{
		\begin{minipage}[t]{0.33\linewidth}
			\centering
			\includegraphics[width=1in]{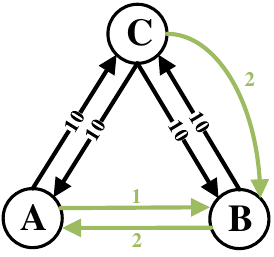}
			\label{fig_PCN2}
		\end{minipage}%
	}%
	\subfloat[A deadlock at \textit{C}.]{
		\begin{minipage}[t]{0.33\linewidth}
			\centering
			\includegraphics[width=1in]{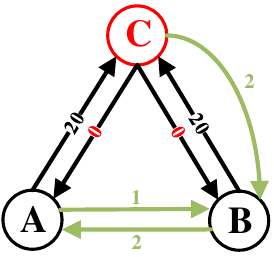}
			\label{fig_PCN3}
		\end{minipage}
	}%
	\centering
	\vspace{-0.2cm}
	\caption{Examples illustrating directional liquidity in a PCN.}
	\vspace{-0.2cm}
\end{figure}

An example payment channel network (PCN) is shown in Fig. \ref{fig_PCN1}, where bidirectional payment channels connect nodes $(A,C)$ and $(C,B)$. Each direction initially holds 10 tokens. When $A$ transfers five tokens to $B$, intermediary $C$ forwards the payment over the two channels and receives a routing fee. A hash timelock contract (HTLC) ensures that $C$ can claim the incoming conditional payment only if it completes the outgoing transfer before the timeout.

\subsection{Transaction Deadlocks in PCNs}\label{deadlock}
To illustrate a local liquidity deadlock, consider Fig. \ref{fig_PCN2}. The offered flows are $A\rightarrow B$ at 1 token/s through $C$, $B\rightarrow A$ at 2 tokens/s through $C$, and an additional $C\rightarrow B$ flow at 2 tokens/s. Each channel initially holds 10 tokens in each direction. On channel $(A,C)$, $C$ forwards 2 tokens/s toward $A$ but receives only 1 token/s from $A$; on channel $(C,B)$, it sends 3 tokens/s toward $B$ (one relayed token plus two local tokens) but receives only 2 tokens/s from $B$. Consequently, $C$ loses one unit of outbound liquidity per second on both channels. Once either outbound balance is exhausted, otherwise feasible payments can no longer be forwarded through $C$, and achieved throughput falls below the offered load, as shown in Fig. \ref{fig_PCN3}. This example motivates rate control that accounts for directional liquidity rather than aggregate channel capacity alone.

\subsection{Trusted Execution Environment}
SHARE uses Intel SGX as its trusted execution environment (TEE). SGX isolates enclave code and data from the host operating system, while remote attestation allows a verifier to check the enclave measurement and platform security state before provisioning secrets or accepting protocol messages.

The evidence is verified through the applicable SGX attestation infrastructure, such as IAS for legacy EPID platforms or DCAP/PCS for ECDSA platforms. The quote authenticates the enclave identity and trusted-computing-base status; it does not automatically authenticate arbitrary application outputs. SHARE therefore binds an enclave-held application verification key to the attested session and uses that key to authenticate subsequent routing and decryption-authorization messages. The parties establish an encrypted channel only after verifying the platform evidence.

\section{Additional Details of System Model}\label{Models}
\subsection{Topology Structure}
\begin{figure}[h]
	\vspace{-0.2cm}
	\centering
	\subfloat[Star-shaped topology.]{
		\begin{minipage}[t]{0.5\linewidth}
			\centering
			\includegraphics[width=1.5in]{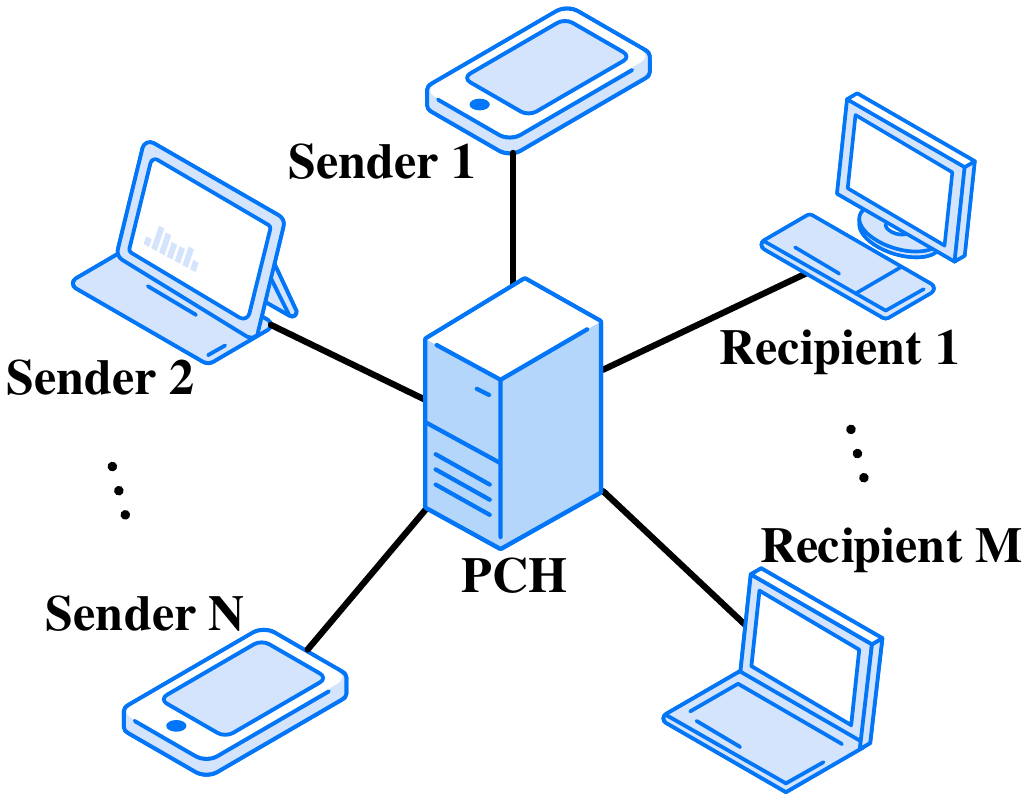}
			\label{PCH1}
		\end{minipage}%
	}%
	\subfloat[Multi-star topology.]{
		\begin{minipage}[t]{0.5\linewidth}
			\centering
			\includegraphics[width=1.5in]{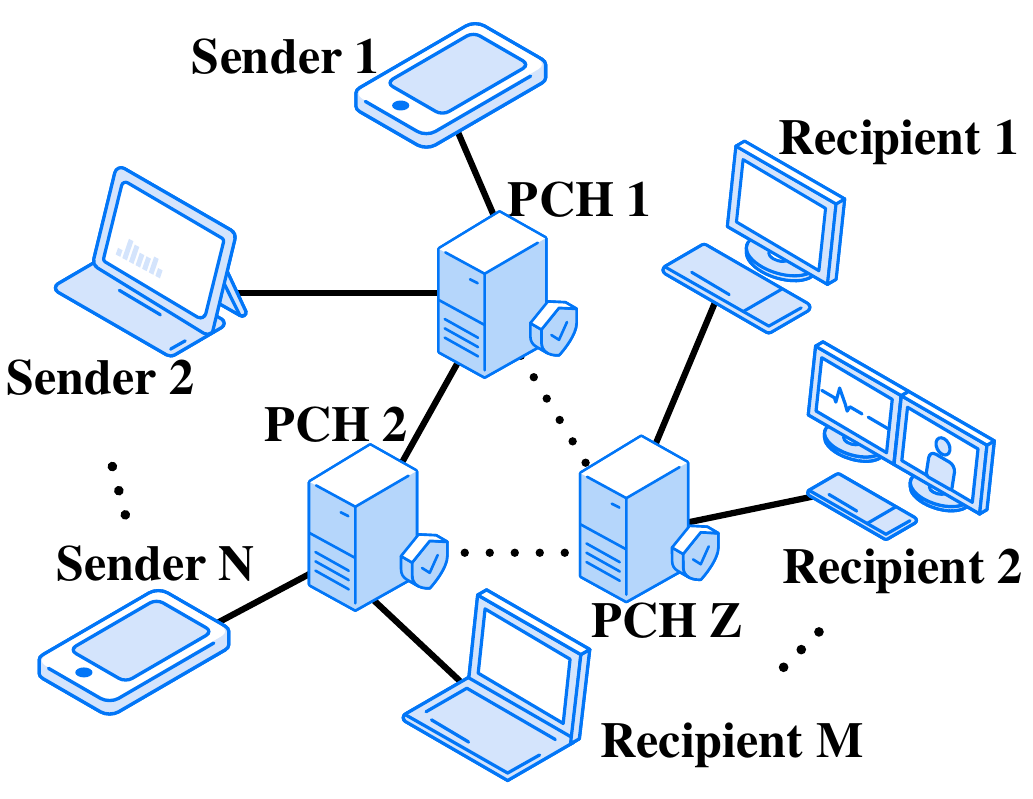}
			\label{PCH2}
		\end{minipage}
	}%
	\centering
	\vspace{-0.2cm}
	\caption{PCN topologies with PCHs.}
	\vspace{-0.2cm}
\end{figure}

Fig. \ref{PCH1} illustrates the star topology commonly used by single-PCH schemes, where clients maintain channels with one intermediary. SHARE generalizes this structure to a multi-star topology in which clients are assigned across multiple TEE-enabled smooth nodes. Fig. \ref{PCH2} shows an example with $N$ senders, $M$ receivers, and $Z$ smooth nodes.

\begin{Def}\label{FedPCN}
(Multi-star PCN topology). A multi-star PCN consists of multiple directly or indirectly interconnected PCHs, and each client conducts transactions through its assigned directly connected PCH.
\end{Def}

\subsection{Trust Model}\label{TM}
SHARE takes an authenticated candidate set as input to the allocation problem. Candidate nomination, governance, collateral, and dispute resolution are external deployment policies and are not modeled by the routing protocol or the UC theorem. A deployment may use voting or stake-based governance to construct this set, but SHARE requires only that every admitted candidate can produce valid platform evidence and that the allocation contract receives authenticated cost inputs.

At the beginning of an allocation epoch, the contract evaluates the authenticated candidate set using the objective in Section \ref{detail_problem}. The resulting active set remains fixed for that epoch and is recomputed when cost inputs or candidate eligibility change. Consensus on the contract output is provided by the underlying blockchain; SHARE does not assume that all candidates independently observe identical request distributions or reach an off-chain unanimous decision.

\begin{figure}[h]
	\vspace{-0.1cm}
	\centering
	\includegraphics[width=3.5in]{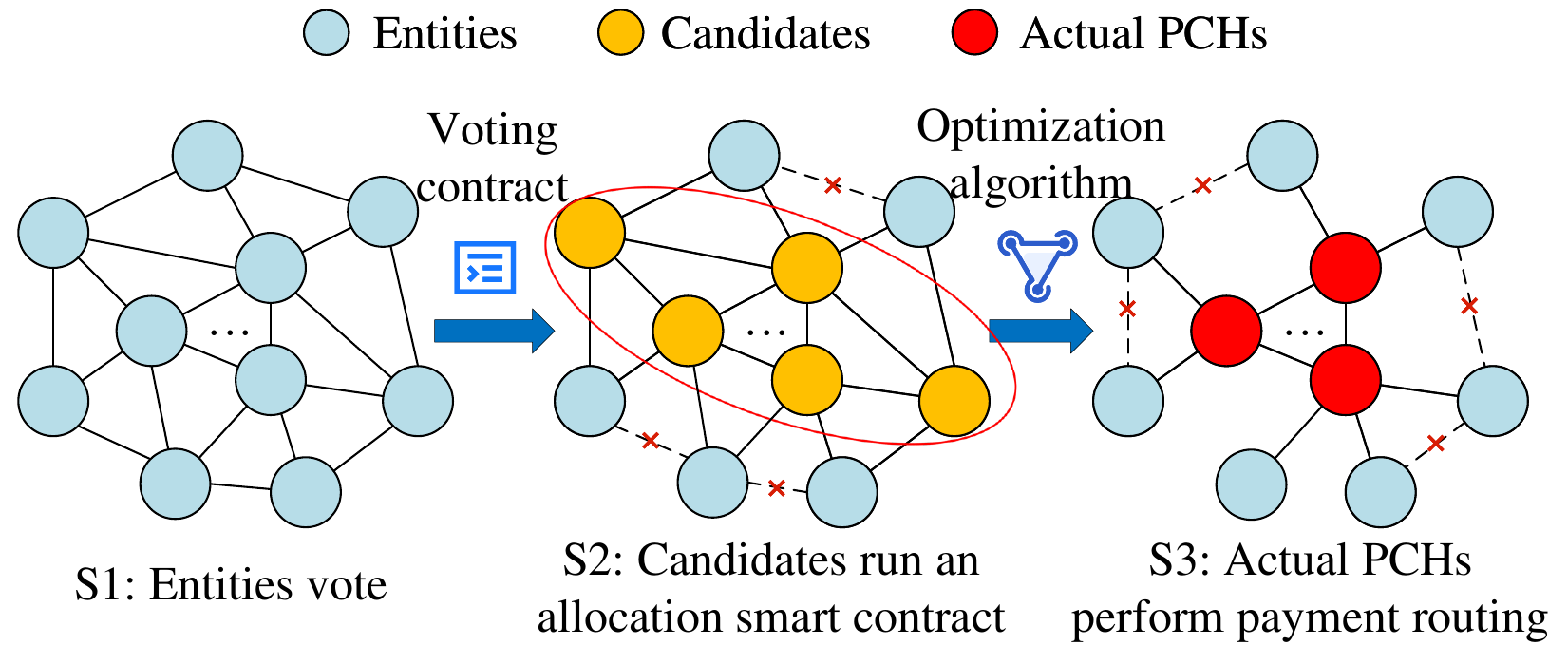}
	\vspace{-0.2cm}
	\caption{Separation between candidate governance and SHARE allocation.}
	\label{trust}
	\vspace{-0.2cm}
\end{figure}

Fig. \ref{trust} illustrates the separation between external candidate governance and SHARE's allocation step. The security analysis does not claim that peers can always detect collusion or that collateral alone deters malicious behavior. Instead, it relies on verified enclave execution, the stated threshold-corruption bound, and authenticated communication. Sensitive transaction values and routing inputs are released only inside authorized enclaves, subject to the leakage defined by $\mathcal{F}_{SHARE}$.

\subsection{Communication Model}
\begin{figure}[h]
	\vspace{-0.2cm}
	\centering
	\includegraphics[width=3.5in]{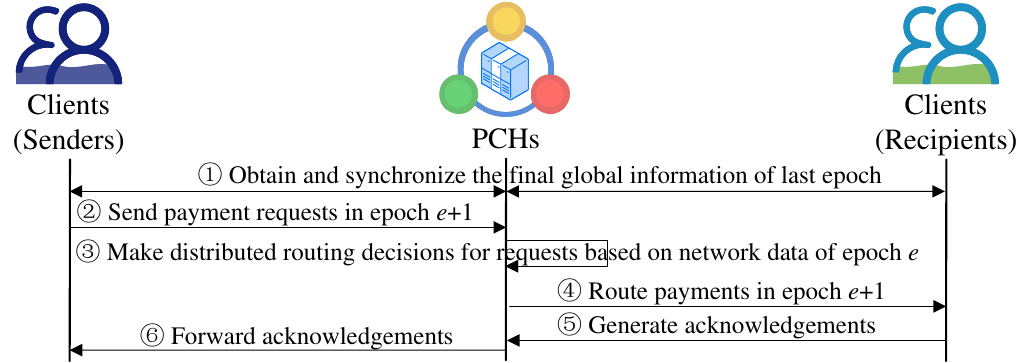}
	\vspace{-0.3cm}
	\caption{Communication process of SHARE in epoch $e+1$.}
	\label{com}
	\vspace{-0.2cm}
\end{figure}

As shown in Fig. \ref{com}, SHARE uses bounded communication epochs. At the beginning of epoch $e+1$, each PCH synchronizes the finalized state from epoch $e$, including topology, channel conditions, and aggregate flow rates. Each PCH then combines this preceding global state with newly received local requests to update its routing decisions. Receiver acknowledgments are returned through the corresponding PCHs. The security proof additionally requires all sensitive messages and threshold-decryption shares to travel over attestation-authenticated channels.

\section{Problem Statement}\label{Sec_P}
\textbf{Multi-PCH allocation problem.} A long-running allocation contract selects active PCHs from the authenticated candidate set and assigns clients to them. The active set remains fixed during an allocation epoch and is recomputed when cost inputs or candidate eligibility change. The model balances client-management, deployment, and inter-PCH synchronization costs; Sections \ref{detail_problem} and \ref{solutions} give the formal objective and solution methods.

Geographically distributed clients experience different communication and routing costs. Activating more PCHs can reduce client-side management cost but increases deployment and synchronization overhead. SHARE therefore seeks a cost-aware deployment rather than assuming either one global hub or one hub per client cluster.

\textbf{Routing problem.} Sender-side source routing becomes increasingly expensive as the network and concurrent workload grow. SHARE delegates routing decisions to multiple PCHs and evaluates networks with up to 3,000 nodes. Inspired by packet switching, it splits a payment into independently routable TUs and controls their rates subject to directional liquidity, in-flight capacity, and \textit{Min-TU}/\textit{Max-TU} constraints. Spider \cite{Sivaraman2020HighTC} demonstrates the utility of multipath rate control; SHARE adapts this principle to attested multi-PCH coordination.

Different intermediate nodes may participate in paths selected for different TUs. Assuming that the underlying PCH construction provides relationship unlinkability, splitting does not intentionally expose additional endpoint linkage beyond the leakage specified by $\mathcal{F}_{SHARE}$. SHARE's UC theorem covers routing confidentiality and execution integrity, not a new proof of relationship unlinkability. Section \ref{protocol} presents the routing protocol.

\section{Security Analysis} \label{Proof}

\subsection{The Real-World Protocol Model \texorpdfstring{$\textbf{Prot}_{SHARE}$}{Prot-SHARE}}
Table \ref{notation} summarizes the notation used in the protocol formalization, which follows Canetti's UC framework \cite{Canetti2001}. We formalize SHARE as $\textbf{Prot}_{SHARE}$ in Algorithm \ref{SHAREProtocol}. The protocol is defined in a hybrid model with two ideal functionalities: (i) the attested-execution functionality $\textcolor{brickred}{\mathcal{G}_{\rm{att}}}$ of Pass \textit{et al.}\footnote{``Formal Abstractions for Attested Execution Secure Processors" (EUROCRYPT 2017); see Algorithm \ref{Gatt}.}, and (ii) the blockchain functionality $\textcolor{brickred}{\mathcal{F}_{\rm{bc}}}$ in Algorithm \ref{FB}.

\begin{table}[h]
	%\vspace{-0.2cm}
	\centering
	\caption{Main notation used in the formal description.}\vspace{-0.2cm}
	\resizebox{1\columnwidth}{!}{
		\begin{tabular}{c l}
			\toprule
			Notation & Description \\
			\midrule
			$\textcolor{brickred}{\mathcal{P}_i}$ & A client as the sender or the recipient \\
			$\textcolor{brickred}{\mathcal{S}_i}$ & A smooth node as the PCH or a member in the KMG \\	
			$\lambda$ & A security parameter \\
			$\mathcal{TE}$ & A robust threshold public-key encryption scheme \\
			$\Sigma$ & A digital signature scheme \\
			\textsf{H} & A hash function \\
			$\iota,t,\tau_{\rm dec}$ & KMG size, corruption bound, and threshold $\tau_{\rm dec}=t+1$ \\
			$\textcolor{colorwheel}{\textsf{tid}} / \textcolor{colorwheel}{\textsf{tuid}}$ & A unique identifier for the transaction/transaction-unit \\
			$(h_{\textsf{tid}},z_{\textsf{tid}})$ & Receiver-generated payment hash and its preimage \\
			$D_i$ & A payment demand for a transaction $i$ \\
			$\textsf{ACK}_{\textcolor{colorwheel}{\textsf{tid}}} / \textsf{ACK}_{\textcolor{colorwheel}{\textsf{tuid}}}$ & An acknowledgment for $\textcolor{colorwheel}{\textsf{tid}} / \textcolor{colorwheel}{\textsf{tuid}}$ \\
			($\textsf{pk}_i,\textsf{vk}_i,\{\textsf{sk}_i^{(q)}\}_{q\in[\iota]}$) & Public key, verification material, and sealed key shares for $i$ \\
			($\textsf{mpk}, \textsf{msk}$) & The attestation public and secret key pair of a processor \\
			\bottomrule
		\end{tabular}%
	}
	\label{notation}%
	%\vspace{-0.3cm}
\end{table}%

\begin{algorithm}[h]
	\DontPrintSemicolon
	\normalem %去除掉下划线
	\caption{\small{TEE Ideal Functionality $\textcolor{brickred}{\mathcal{G}_{\rm{att}}}[\Sigma, \textsf{reg}]$}} \label{Gatt}
	\scriptsize{
		\tcp{\scriptsize{initialization}}
		\textcolor{applegreen}{\textbf{On initialize}}: ($\textsf{mpk}, \textsf{msk}
		$) $:= \Sigma.\textsf{KeyGen}$($1^{\lambda}$), $T = \emptyset$
		
		\tcp{\scriptsize{public query interface}}
		\textcolor{colorwheel}{\textbf{On receive}} (\textcolor{mediumorchid}{``getpk()"}) from some $\textcolor{brickred}{\mathcal{P}}$: send \textsf{mpk} to $\textcolor{brickred}{\mathcal{P}}$ 
		
		\tcp{\rule[2.5pt]{2cm}{0.05em} \scriptsize{Enclave operations} \rule[2.5pt]{2cm}{0.05em}}
		
		\tcp{\scriptsize{local interface $-$ install an enclave}}
		\textcolor{colorwheel}{\textbf{On receive}} (\textcolor{mediumorchid}{``install"}$,$ $idx$$,$ \textsf{prog}) from some $\textcolor{brickred}{\mathcal{P}} \in \textsf{reg}:$
		
		\quad if $\textcolor{brickred}{\mathcal{P}}$ is honest, assert $idx = sid$; generate nonce $eid \in \{0, 1\}^{\lambda}$,\\ 
		\quad store $T[eid, \textcolor{brickred}{\mathcal{P}}] := $($idx, \textsf{prog}, \bm{0}$), send $eid$ to $\textcolor{brickred}{\mathcal{P}}$
		
		\tcp{\scriptsize{local interface $-$ resume an enclave}}
		\textcolor{colorwheel}{\textbf{On receive}} (\textcolor{mediumorchid}{``resume"}$,$ $eid$$,$ \textsf{inp}) from some $\textcolor{brickred}{\mathcal{P}} \in \textsf{reg}:$
		
		\quad let ($idx, \textsf{prog}, \textsf{mem}$) $ := T[eid, \textcolor{brickred}{\mathcal{P}}]$; abort if not found
		
		\quad let ($\textsf{outp}, \textsf{mem}$) $ := \textsf{prog}$($\textsf{inp}, \textsf{mem}$), \\
		\quad update $T[eid, \textcolor{brickred}{\mathcal{P}}] := $ ($idx, \textsf{prog}, \textsf{mem}$)
		
		\quad let $\sigma := \Sigma.\textsf{Sig}_{\textsf{msk}}$($idx, eid, \textsf{prog}, \textsf{outp}$), send ($\textsf{outp}, \sigma$) to $\textcolor{brickred}{\mathcal{P}}$
	}
\end{algorithm}

\begin{algorithm}[h]
	\DontPrintSemicolon
	\normalem %去除掉下划线
	\caption{\small{Blockchain Ideal Functionality $\textcolor{brickred}{\mathcal{F}_{\rm{bc}}}[\textsf{succ}]$}} \label{FB}
	\scriptsize{
		Parameter: successor relationship \textsf{succ}:$\{0,1\}^*\times\{0,1\}^*\rightarrow\{0,1\}$\\
		\tcp{\scriptsize{initialization}}
		\textcolor{colorwheel}{\textbf{On receive}} (\textcolor{mediumorchid}{``init"}): $\textsf{Storage} := \emptyset$\\
		\tcp{\scriptsize{public query interface}}
		\textcolor{colorwheel}{\textbf{On receive}} (\textcolor{mediumorchid}{``read"}, \textcolor{colorwheel}{\textsf{tid}}): output \textsf{Storage}[\textcolor{colorwheel}{\textsf{tid}}], or $\bot$ if not found\\
		\tcp{\scriptsize{public append interface}}
		\textcolor{colorwheel}{\textbf{On receive}} (\textcolor{mediumorchid}{``append"}, \textcolor{colorwheel}{\textsf{tid}}, \textsf{inp}) from $\textcolor{brickred}{\mathcal{P}}$:\\
		\quad let $\textsf{v} := \textsf{Storage}[\textcolor{colorwheel}{\textsf{tid}}]$, set to $\bot$ if not found \\
		\quad if $\textsf{succ}(\textsf{v}, \textsf{inp}) = 1$ then\\ 
		\qquad $\textsf{Storage}[\textcolor{colorwheel}{\textsf{tid}}] := (\textsf{inp}, \textcolor{brickred}{\mathcal{P}})$, output (\textcolor{mediumorchid}{``success"}$, \textcolor{colorwheel}{\textsf{tid}}$)\\
		\quad else output (\textcolor{mediumorchid}{``failure"}$, \textcolor{colorwheel}{\textsf{tid}}$)
	}
\end{algorithm}

$\textcolor{brickred}{\mathcal{G}_{\rm{att}}}$ abstracts a general-purpose secure processor. As shown in Algorithm \ref{Gatt}, initialization generates an attestation key pair $(\textsf{mpk},\textsf{msk})$. The processor keeps $\textsf{msk}$ private and exposes $\textsf{mpk}$ through ``\verb|getpk()|". A party invokes ``\verb|install|" to create an enclave, load $\textsf{prog}$, and obtain a fresh enclave identifier $eid$. On ``\verb|resume|", $\textcolor{brickred}{\mathcal{G}_{\rm{att}}}$ executes $\textsf{prog}$ and signs the output and its execution metadata under $\textsf{msk}$.

\begin{algorithm*}[h]
	\DontPrintSemicolon
	\normalem %去除掉下划线
	\caption{\small{SHARE's Formal Protocol in Real-World Model $\textbf{Prot}_{SHARE}(\lambda, \mathcal{TE}, \Sigma, \iota,t,\tau_{\rm dec}, \{\textcolor{brickred}{\mathcal{P}_i}\}_{i \in [\mathbb{V}_{\rm CLI}]}, \{\textcolor{brickred}{\mathcal{S}_i}\}_{i \in [\mathbb{V}_{\rm SN}]})$}} \label{SHAREProtocol}
	%\KwInitially{$\textsf{funds\_r} := \emptyset$}
	%\KwOutput{$X_{z}^s$ (or equivalently $Y_{z}^s$)}
	
	\begin{multicols}{3}
		\scriptsize{
			
			\underline{\footnotesize{Clients $\textcolor{brickred}{\mathcal{P}_i}$:}} \\
			\tcp{\scriptsize{Part 1: payment initialization}}
			\textcolor{colorwheel}{\textbf{On receive}} (\textcolor{mediumorchid}{``init"}$, \textsf{pay}_{\rm{req}},h_{\textsf{tid}}$) from environment $\textcolor{brickred}{\mathcal{Z}}$: \label{A3l2}
			
			\quad send (\textcolor{mediumorchid}{``init"}$, \textsf{pay}_{\rm{req}},h_{\textsf{tid}}$) to $\textcolor{brickred}{\mathcal{S}_i}$, 
			
			\quad wait for ($\textcolor{colorwheel}{\textsf{tid}}, \textsf{pk}_{\textcolor{colorwheel}{\textsf{tid}}}, \textsf{mpk}_i, \textsf{state}_{\textcolor{colorwheel}{\textsf{tid}}}$)
			
			\quad \textbf{return} ($\textcolor{colorwheel}{\textsf{tid}}, \textsf{pk}_{\textcolor{colorwheel}{\textsf{tid}}}, \textsf{mpk}_i, \textsf{state}_{\textcolor{colorwheel}{\textsf{tid}}}$) \label{A3l5}\\
			
			\tcp{\scriptsize{Part 2: payment processing}}
			\textcolor{colorwheel}{\textbf{On receive}} (\textcolor{mediumorchid}{``pay\_t"}$, \textcolor{colorwheel}{\textsf{tid}}, D_{\textcolor{colorwheel}{\textsf{tid}}}$) from environment $\textcolor{brickred}{\mathcal{Z}}$: \label{A3l6}
			
			\quad parse $D_{\textcolor{colorwheel}{\textsf{tid}}}$ as ($P_s, P_r, \textsf{val}_{\textcolor{colorwheel}{\textsf{tid}}},h_{\textsf{tid}}$) \tcp{\scriptsize{Here $\textcolor{brickred}{\mathcal{P}_i}$ is $P_s$}}
			
			\quad $\textsf{inp} := \mathcal{TE}.\textsf{Enc}$($\textsf{pk}_{\textcolor{colorwheel}{\textsf{tid}}}, D_{\textcolor{colorwheel}{\textsf{tid}}}$)
			
			\quad \tcp{\scriptsize{Let $\textcolor{brickred}{\mathcal{S}_{i}}$ be assigned to sender $P_s$}}
			\quad lock $\$\textsf{val}_{\textcolor{colorwheel}{\textsf{tid}}}$ to $\textcolor{brickred}{\mathcal{S}_{i}}$ under $h_{\textsf{tid}}$ and an HTLC expiry, \\
			\quad send (\textcolor{mediumorchid}{``pay\_t"}$, \textcolor{colorwheel}{\textsf{tid}}, \textsf{inp}$) to $\textcolor{brickred}{\mathcal{S}_i}$, wait for ($\sigma_{\textcolor{colorwheel}{\textsf{tid}}}, \tiny{\textsf{ACK}_{\textcolor{colorwheel}{\textsf{tid}}}}$)\\
			\quad assert $\tiny{\textsf{ACK}_{\textcolor{colorwheel}{\textsf{tid}}}}, $ \textbf{return} $\top$ \label{A3l12}
			
			\tcp{\scriptsize{Part 3: payment acknowledgment}}
			\textcolor{colorwheel}{\textbf{On receive}} (\textcolor{mediumorchid}{``release"}$, \textcolor{colorwheel}{\textsf{tid}}, D_{\textcolor{colorwheel}{\textsf{tid}}}$) from $\textcolor{brickred}{\mathcal{S}_i}$: \label{A3l13}
			
			\quad parse $D_{\textcolor{colorwheel}{\textsf{tid}}}$ as ($P_s, P_r, \textsf{val}_{\textcolor{colorwheel}{\textsf{tid}}},h_{\textsf{tid}}$) \tcp{\scriptsize{Here $\textcolor{brickred}{\mathcal{P}_i}$ is $P_r$}}
			\quad assert all $K$ TU locks are committed and $\textsf{H}(z_{\textsf{tid}})=h_{\textsf{tid}}$; send $z_{\textsf{tid}}$ to $\textcolor{brickred}{\mathcal{S}_{i}}$\\
			\quad after settlement, send $\textsf{ACK}_{\textsf{tid}}$ to $\textcolor{brickred}{\mathcal{S}_{i}}$ and \textbf{return} $\top$ \label{A3l16}
			%\end{algorithm*}
			
			%\SetNlSty{textbf}{}{} %换页伪代码行号风格
			%\begin{algorithm*}[t]
			%	\setcounter{AlgoLine}{16}
			%	\SetAlgoVlined
			\underline{\footnotesize{Smooth nodes $\textcolor{brickred}{\mathcal{S}_i}$:}} \\ 
			
			\tcp{\scriptsize{Part 1: payment initialization}}
			\textcolor{applegreen}{\textbf{On initialize}}: $\textsf{eid}_i\leftarrow\bot$, $\textsf{mpk}_i\leftarrow\bot$ \label{A3l18}
			
			\textcolor{colorwheel}{\textbf{On receive}} (\textcolor{mediumorchid}{``install"}$,$ \textsf{prog}) from environment $\textcolor{brickred}{\mathcal{Z}}$:
			
			\quad send (\textcolor{mediumorchid}{``install"}$,$ i,\textsf{prog}) to $\textcolor{brickred}{\mathcal{G}_{\rm{att}}}$, wait for $\textsf{eid}_i$; set $\textsf{mpk}_i\leftarrow\mathcal{G}_{\rm att}.\textsf{getpk}()$\\
			\quad \textbf{return} $(\textsf{eid}_i,\textsf{mpk}_i)$; the secret attestation key never leaves $\mathcal{G}_{\rm att}$ \label{A3l21}
			
			\tcp{\scriptsize{KMG setup: $\iota\geq2t+1$ and $\tau_{\rm dec}=t+1$}}
			\textcolor{applegreen}{\textbf{Offline refill}}($B$):
			\quad\textbf{for} $u=1$ \textbf{to} $B$ \textbf{do} jointly run
			$(\textsf{pk}_{u},\textsf{vk}_{u},\{\textsf{sk}_{u}^{(q)}\}_{q\in[\iota]})
			\leftarrow\mathcal{TE}.\textsf{DKG}(1^{\lambda},\iota,\tau_{\rm dec})$\;
			\quad each member $q$ seals $\textsf{sk}_{u}^{(q)}$; add $(u,\textsf{pk}_{u},\textsf{vk}_{u},\textsf{unused})$ to $\mathsf{Pool}$\;
			
			\textcolor{colorwheel}{\textbf{On receive}} (\textcolor{mediumorchid}{``Allocate"}$,$ $x$) from $\textcolor{brickred}{\mathcal{S}_{i'}}$: \label{A3l22}
			\quad atomically select an unused $u\in\mathsf{Pool}$; abort if none exists\;
			\quad set $\mathsf{Bind}[x]\leftarrow u$, mark $u$ allocated, and alias $(\textsf{pk}_{x},\textsf{vk}_{x},\textsf{sk}_{x}^{(q)})\leftarrow(\textsf{pk}_{u},\textsf{vk}_{u},\textsf{sk}_{u}^{(q)})$\;
			\quad send $(x,\textsf{pk}_{x},\textsf{vk}_{x})$ to $\textcolor{brickred}{\mathcal{S}_{i'}}$ \label{A3l24}
			
			\textcolor{colorwheel}{\textbf{On receive}} (\textcolor{mediumorchid}{``PDec"}$,$ $x,\textsf{ct}_{x},\textsf{eid}_{i'},\textsf{auth}_{x},\pi_x^{\rm dec},\textsf{mpk}_{i'}$) over an attested channel: \label{A3l25}
			\quad assert $\mathsf{Bind}[x]$ is allocated, $\textsf{auth}_{x}=(\textsf{authorize\_dec},x,\textsf{H}(\textsf{ct}_{x}))$, and\\
			\quad $\Sigma.\textsf{Vrfy}_{\textsf{mpk}_{i'}}((i',\textsf{eid}_{i'},\textsf{prog},\textsf{auth}_{x}),\pi_x^{\rm dec})=1$\\
			\quad each responsive KMG member $q$ computes $\delta_{x}^{(q)}\leftarrow\mathcal{TE}.\textsf{PDec}(\textsf{pk}_{x},\textsf{sk}_{x}^{(q)},\textsf{ct}_{x})$\\
			\quad deliver $(q,\delta_{x}^{(q)})$ only to enclave $\textsf{eid}_{i'}$ over that channel; never release $\textsf{sk}_{x}^{(q)}$ \label{A3l27}
			
			\textcolor{colorwheel}{\textbf{On receive}} (\textcolor{mediumorchid}{``init"}$,$ $\textsf{pay}_{\rm{req}},h_{\textsf{tid}}$) from $\textcolor{brickred}{\mathcal{P}_i}$: \label{A3l28}
			
			\quad \textcolor{colorwheel}{\textsf{tid}} $\leftarrow\$\{0,1\}^{\lambda}$
			
			\quad send (\textcolor{mediumorchid}{``Allocate"}$,$ \textcolor{colorwheel}{\textsf{tid}}) to KMG, wait for ($\textsf{pk}_{\textcolor{colorwheel}{\textsf{tid}}},\textsf{vk}_{\textcolor{colorwheel}{\textsf{tid}}}$)
			
			\quad $\theta_{\textcolor{colorwheel}{\textsf{tid}}} := \bot$; $\textsf{state}_{\textcolor{colorwheel}{\textsf{tid}}}:= $ ($\textcolor{colorwheel}{\textsf{tid}}, \theta_{\textcolor{colorwheel}{\textsf{tid}}},h_{\textsf{tid}}$)
			
			\quad send ($\textcolor{colorwheel}{\textsf{tid}}, \textsf{pk}_{\textcolor{colorwheel}{\textsf{tid}}}, \textsf{mpk}_i, \textsf{state}_{\textcolor{colorwheel}{\textsf{tid}}}$) to $\textcolor{brickred}{\mathcal{P}_i}$ \label{A3l33}
			
			\tcp{\scriptsize{Part 2: payment processing}}
			\textcolor{colorwheel}{\textbf{On receive}} (\textcolor{mediumorchid}{``pay\_t"}$, \textcolor{colorwheel}{\textsf{tid}}, \textsf{inp}$) from $\textcolor{brickred}{\mathcal{P}_i}$: \label{A3l34}
			
			\quad send $(\textcolor{mediumorchid}{``resume"}$,$\textsf{eid}_i,(\textsf{authorize\_dec},\textcolor{colorwheel}{\textsf{tid}},\textsf{H}(\textsf{inp})))$ to $\mathcal{G}_{\rm att}$; receive $(\textsf{auth}_{\textsf{tid}},\pi_{\textsf{tid}}^{\rm dec})$\\
			\quad send $(\textcolor{mediumorchid}{``PDec"}$,$ \textcolor{colorwheel}{\textsf{tid}},\textsf{inp},\textsf{eid}_i,\textsf{auth}_{\textsf{tid}},\pi_{\textsf{tid}}^{\rm dec},\textsf{mpk}_i)$ to KMG; collect shares inside $\textsf{eid}_i$
			
			\quad $Q_{\textcolor{colorwheel}{\textsf{tid}}}:=\{q:\mathcal{TE}.\textsf{VerifyShare}(\textsf{pk}_{\textcolor{colorwheel}{\textsf{tid}}},\textsf{vk}_{\textcolor{colorwheel}{\textsf{tid}}},\textsf{inp},\delta_{\textcolor{colorwheel}{\textsf{tid}}}^{(q)})=1\}$; abort if $|Q_{\textcolor{colorwheel}{\textsf{tid}}}|<\tau_{\rm dec}$
			
			\quad $D_{\textcolor{colorwheel}{\textsf{tid}}}:=\mathcal{TE}.\textsf{Combine}(\textsf{pk}_{\textcolor{colorwheel}{\textsf{tid}}},\textsf{vk}_{\textcolor{colorwheel}{\textsf{tid}}},\textsf{inp},\{\delta_{\textcolor{colorwheel}{\textsf{tid}}}^{(q)}\}_{q\in Q_{\textcolor{colorwheel}{\textsf{tid}}}})$
			
			\quad parse $D_{\textcolor{colorwheel}{\textsf{tid}}}$ as ($P_s, P_r, \textsf{val}_{\textcolor{colorwheel}{\textsf{tid}}},h_{\textsf{tid}}$)
			
			\quad\textbf{for} $k = 1$ \textbf{\textit{to}} $K$ \textbf{do}\\
			\qquad $\textcolor{colorwheel}{\textsf{tuid}}_k\leftarrow\$\{0,1\}^{\lambda}$ \\
			\quad split $D_{\textcolor{colorwheel}{\textsf{tid}}}$ into $\{D_{\textcolor{colorwheel}{\textsf{tuid}}_k}\}_{1\leq k\leq K}$; each TU carries $(\textsf{tid},k,K,h_{\textsf{tid}})$ \tcp{\scriptsize{Algorithm 2}}
			\quad\textbf{for} $k = 1$ \textbf{\textit{to}} $K$ \textbf{do}\\
			\qquad $\theta_{\textcolor{colorwheel}{\textsf{tuid}}_k}:=\bot$; $\textsf{state}_{\textcolor{colorwheel}{\textsf{tuid}}_k}:=(\textcolor{colorwheel}{\textsf{tuid}}_k,\theta_{\textcolor{colorwheel}{\textsf{tuid}}_k})$ \\
			\qquad parse $D_{\textcolor{colorwheel}{\textsf{tuid}}_k}$ as ($P_s,P_r,\textsf{val}_{\textcolor{colorwheel}{\textsf{tuid}}_k},\textsf{tid},k,K,h_{\textsf{tid}}$)\\
			\qquad send (\textcolor{mediumorchid}{``Allocate"}$,$ $\textcolor{colorwheel}{\textsf{tuid}}_k$) to KMG, wait for $(\textsf{pk}_{\textcolor{colorwheel}{\textsf{tuid}}_k},\textsf{vk}_{\textcolor{colorwheel}{\textsf{tuid}}_k})$\\
			\qquad $\textsf{inp}_{{\rm tu},k}:=\mathcal{TE}.\textsf{Enc}(\textsf{pk}_{\textcolor{colorwheel}{\textsf{tuid}}_k},D_{\textcolor{colorwheel}{\textsf{tuid}}_k})$ \\
			\quad\textbf{in parallel for} $k=1$ \textbf{to} $K$ \textbf{do}\\
			\qquad lock $\$\textsf{val}_{\textcolor{colorwheel}{\textsf{tuid}}_k}$ under $h_{\textsf{tid}}$ and send (\textcolor{mediumorchid}{``pay\_tu"}$,$ $\textcolor{colorwheel}{\textsf{tuid}}_k,\textsf{pk}_{\textcolor{colorwheel}{\textsf{tuid}}_k},\textsf{vk}_{\textcolor{colorwheel}{\textsf{tuid}}_k},\textsf{inp}_{{\rm tu},k}$) to $\textcolor{brickred}{\mathcal{S}_{j}}$\\
			\quad\textbf{for} $k=1$ \textbf{to} $K$ \textbf{do} wait for $\textsf{READY}_{\textcolor{colorwheel}{\textsf{tuid}}_k}$; on failure or timeout, reveal no preimage and let pending HTLCs time out\\
			\quad\textbf{for} $k=1$ \textbf{to} $K$ \textbf{do} wait for $\textsf{ACK}_{\textcolor{colorwheel}{\textsf{tuid}}_k}$ after preimage fulfillment; set $\theta_{\textcolor{colorwheel}{\textsf{tuid}}_k}\leftarrow\top$ \label{A3l50}\\
			\quad $\theta_{\textcolor{colorwheel}{\textsf{tid}}}:=\bigwedge_{k=1}^{K}\theta_{\textcolor{colorwheel}{\textsf{tuid}}_k}$; update $\textsf{state}_{\textcolor{colorwheel}{\textsf{tid}}}$
			
			\quad assert $\theta_{\textcolor{colorwheel}{\textsf{tid}}}$; send $(\textcolor{mediumorchid}{``resume"}$,$\textsf{eid}_i,(\textsf{finalize},\textsf{state}_{\textcolor{colorwheel}{\textsf{tid}}}))$ to $\mathcal{G}_{\rm att}$\\
			\quad wait for $(\textsf{outp}_{\textsf{tid}},\sigma_{\textcolor{colorwheel}{\textsf{tid}}})$; assert $\textsf{outp}_{\textsf{tid}}=(\textsf{finalized},\textsf{state}_{\textcolor{colorwheel}{\textsf{tid}}})$
			
			\quad wait for $\textsf{ACK}_{\textcolor{colorwheel}{\textsf{tid}}}$ from $\textcolor{brickred}{\mathcal{S}_{j}}$ over their attested secure channel
			
			\quad assert $\tiny{\textsf{ACK}_{\textcolor{colorwheel}{\textsf{tid}}}}$, send 
			($\sigma_{\textcolor{colorwheel}{\textsf{tid}}}, \tiny{\textsf{ACK}_{\textcolor{colorwheel}{\textsf{tid}}}}$) to $\textcolor{brickred}{\mathcal{P}_i}$, \textbf{return} $\top$ \label{A3l54}
			
			\textcolor{colorwheel}{\textbf{On receive}} (\textcolor{mediumorchid}{``pay\_tu"}$,$ $\textcolor{colorwheel}{\textsf{tuid}}_k,\textsf{pk}_{\textcolor{colorwheel}{\textsf{tuid}}_k},\textsf{vk}_{\textcolor{colorwheel}{\textsf{tuid}}_k},\textsf{inp}_{{\rm tu},k}$) from $\textcolor{brickred}{\mathcal{S}_{i'}}$: \label{A3l55}
			
			\quad send $(\textcolor{mediumorchid}{``resume"}$,$\textsf{eid}_i,(\textsf{authorize\_dec},\textcolor{colorwheel}{\textsf{tuid}}_k,\textsf{H}(\textsf{inp}_{{\rm tu},k})))$ to $\mathcal{G}_{\rm att}$; receive $(\textsf{auth}_k,\pi_k^{\rm dec})$\\
			\quad send $(\textcolor{mediumorchid}{``PDec"}$,$ \textcolor{colorwheel}{\textsf{tuid}}_k,\textsf{inp}_{{\rm tu},k},\textsf{eid}_i,\textsf{auth}_k,\pi_k^{\rm dec},\textsf{mpk}_i)$ to KMG; collect shares inside $\textsf{eid}_i$
			
			\quad $Q_k:=\{q:\mathcal{TE}.\textsf{VerifyShare}(\textsf{pk}_{\textcolor{colorwheel}{\textsf{tuid}}_k},\textsf{vk}_{\textcolor{colorwheel}{\textsf{tuid}}_k},\textsf{inp}_{{\rm tu},k},\delta_{\textcolor{colorwheel}{\textsf{tuid}}_k}^{(q)})=1\}$; abort if $|Q_k|<\tau_{\rm dec}$
			
			\quad $D_{\textcolor{colorwheel}{\textsf{tuid}}_k}:=\mathcal{TE}.\textsf{Combine}(\textsf{pk}_{\textcolor{colorwheel}{\textsf{tuid}}_k},\textsf{vk}_{\textcolor{colorwheel}{\textsf{tuid}}_k},\textsf{inp}_{{\rm tu},k},\{\delta_{\textcolor{colorwheel}{\textsf{tuid}}_k}^{(q)}\}_{q\in Q_k})$
			
			\quad parse $D_{\textcolor{colorwheel}{\textsf{tuid}}_k}$ as ($P_s,P_r,\textsf{val}_{\textcolor{colorwheel}{\textsf{tuid}}_k},\textsf{tid},k,K,h_{\textsf{tid}}$)
			
			\quad assert $\textsf{HTLC.Locked}(\textsf{tuid}_k,\textsf{val}_{\textsf{tuid}_k},h_{\textsf{tid}})$; store $D_{\textsf{tuid}_k}$ as pending
			
			\quad send $\textsf{READY}_{\textcolor{colorwheel}{\textsf{tuid}}_k}$ to $\textcolor{brickred}{\mathcal{S}_{i'}}$ \label{A3l59}
			
			\tcp{\scriptsize{Part 3: complete-set acceptance and acknowledgment}}
			\textcolor{colorwheel}{\textbf{When}} all $K$ pending TUs for $\textsf{tid}$ are valid and locked: \label{A3l60}
			\quad send (\textcolor{mediumorchid}{``release"}$,\textsf{tid},D_{\textsf{tid}}$) to $P_r$; receive $z_{\textsf{tid}}$ and assert $\textsf{H}(z_{\textsf{tid}})=h_{\textsf{tid}}$\\
			\quad issue fulfillment for the entire HTLC set with $z_{\textsf{tid}}$; otherwise reveal no preimage and let pending locks time out\\
			\quad for each fulfilled TU, send $\textsf{ACK}_{\textsf{tuid}_k}$ to $\textcolor{brickred}{\mathcal{S}_{i'}}$; wait for $\textsf{ACK}_{\textsf{tid}}$ from $P_r$\\
			\quad send $\textsf{ACK}_{\textsf{tid}}$ to $\textcolor{brickred}{\mathcal{S}_{i'}}$ and \textbf{return} $\top$ \label{A3l65}
			
		}
	\end{multicols}
\end{algorithm*}

The blockchain ideal functionality $\textcolor{brickred}{\mathcal{F}_{\rm{bc}}}$ defines a general-purpose blockchain protocol that models an append-only ledger. As shown in Algorithm \ref{FB}, $\textcolor{brickred}{\mathcal{F}_{\rm{bc}}}$ can append blockchain data associated with a transaction \textcolor{colorwheel}{\textsf{tid}} to an internal \textsf{Storage} via the ``\verb|append|" command. The parameter \textsf{succ} is a function that models the notion of the appended transaction validity. The basic operations of the payment channel, e.g., initialization, creation, and settlement, invoke the functionalities in $\textcolor{brickred}{\mathcal{F}_{\rm{bc}}}$. Due to page limits, we omit these descriptions in $\textbf{Prot}_{SHARE}$.

We emphasize that SHARE's security is designed to be independent of specific blockchain and TEE instances, as long as they provide the functionalities required by $\textcolor{brickred}{\mathcal{F}_{\rm{bc}}}$ and $\textcolor{brickred}{\mathcal{G}_{\rm{att}}}$.

Following the workflow described in \S \ref{sec_SM}, we formalize $\textbf{Prot}_{SHARE}$ (Algorithm \ref{SHAREProtocol}) in three parts:

%\vspace{-0.2cm}
\begin{algorithm*}[h]
	\DontPrintSemicolon
	\normalem %去除掉下划线
	\caption{\small{SHARE's Ideal Functionality $\textcolor{brickred}{\mathcal{F}_{SHARE}}(\lambda,\ell,\iota,t,\tau_{\rm dec},B_{\rm pool},\{\textcolor{brickred}{\mathcal{P}_i}\}_{i\in[\mathbb{V}]})$}} \label{F-SHARE}
	\begin{multicols}{2}
		\scriptsize{
			Parameters: $\ell:\{0,1\}^*\rightarrow\{0,1\}^*$, where $\ell(m)$ includes $|m|$ and permitted metadata; $\iota\geq2t+1$; $\tau_{\rm dec}=t+1$; and pool size $B_{\rm pool}$\\
			\textcolor{applegreen}{\textbf{On initialize}}: $\mathsf{Pool}\leftarrow[B_{\rm pool}]$, $\mathsf{Bind}\leftarrow\emptyset$, $\mathsf{State}\leftarrow\emptyset$\\
			\tcp{At most $t$ KMG members may be statically corrupted; decryption requires $\tau_{\rm dec}$ responsive members}
			\tcp{\scriptsize{Payment initialization}}
			\textcolor{colorwheel}{\textbf{On receive}} (\textcolor{mediumorchid}{``init"}$, \textsf{pay}_{\rm{req}},h_{\textsf{tid}}$) from $\textcolor{brickred}{\mathcal{P}_i}$ for some $i \in [\mathbb{V}_{\rm CLI}]$:\\
			\quad wait if $\mathsf{Pool}=\emptyset$; otherwise remove an unused entry $u$ from $\mathsf{Pool}$\\
			\quad $\textcolor{colorwheel}{\textsf{tid}} \leftarrow \$\{0,1\}^{\lambda}$\\
			\quad bind $u$ to $\textsf{tid}$ and obtain $(\textsf{pk}_{\textsf{tid}},\textsf{vk}_{\textsf{tid}})\leftarrow\textsf{AllocateKey}(u,\textsf{tid})$\\
			\quad notify $\textcolor{brickred}{\mathcal{A}}$ of (\textcolor{mediumorchid}{``init"}$,\textcolor{brickred}{\mathcal{P}_i},\ell(\textsf{pay}_{\rm req}),h_{\textsf{tid}},\textsf{tid}$); block until $\textcolor{brickred}{\mathcal{A}}$ replies\\
			\quad $\mathsf{State}[\textsf{tid}]\leftarrow(\textsf{tid},\bot,h_{\textsf{tid}})$\\
			\quad send private delayed output $(\textsf{tid},\textsf{pk}_{\textsf{tid}},\textsf{mpk}_i,\mathsf{State}[\textsf{tid}])$ to $\textcolor{brickred}{\mathcal{P}_i}$
			
			\tcp{\scriptsize{Payment processing}}
			\textcolor{colorwheel}{\textbf{On receive}} (\textcolor{mediumorchid}{``pay\_t"}$, \textcolor{colorwheel}{\textsf{tid}}, D_{\textcolor{colorwheel}{\textsf{tid}}}$) from $\textcolor{brickred}{\mathcal{P}_i}$ for some $i \in [\mathbb{V}_{\rm CLI}]$:\\
			\quad parse $D_{\textsf{tid}}$ as $(P_s,P_r,\textsf{val}_{\textsf{tid}},h_{\textsf{tid}})$; require the sender's conditional transfer under $h_{\textsf{tid}}$\\
			\quad notify $\textcolor{brickred}{\mathcal{A}}$ of (\textcolor{mediumorchid}{``pay\_t"}$,\textsf{tid},\textcolor{brickred}{\mathcal{P}_i},\ell(D_{\textsf{tid}})$)\\
			\quad wait until $|\mathsf{Pool}|\geq K$ and at least $\tau_{\rm dec}$ KMG members are responsive\\
			\quad\textbf{for} $k=1$ \textbf{to} $K$ \textbf{do}: sample $\textsf{tuid}_k\leftarrow\$\{0,1\}^{\lambda}$ and allocate one pool entry\\
			\quad split $D_{\textsf{tid}}$ into $\{D_{\textsf{tuid}_k}\}_{k=1}^{K}$ carrying the common $h_{\textsf{tid}}$; set $\mathsf{State}[\textsf{tuid}_k]\leftarrow(\textsf{tuid}_k,\bot)$\\
			\quad\textbf{in parallel for} $k=1$ \textbf{to} $K$ \textbf{do} notify $\textcolor{brickred}{\mathcal{A}}$ of (\textcolor{mediumorchid}{``pay\_tu"}$,\textsf{tuid}_k,\ell(D_{\textsf{tuid}_k})$)\\
			\quad if any TU lock fails or expires before release, reveal no preimage, let each pending HTLC resolve by timeout, and output $\textsf{FAIL}$\\
			\quad otherwise request $z_{\textsf{tid}}$ privately from $P_r$; require $\textsf{H}(z_{\textsf{tid}})=h_{\textsf{tid}}$ and set $\mathsf{Accept}[\textsf{tid}]\leftarrow\top$\\
			\quad invoke $\textsf{Fulfill}(\textsf{tuid}_k,z_{\textsf{tid}})$ for every $k$ through the HTLC interface\\
			\quad notify $\textcolor{brickred}{\mathcal{A}}$ of (\textcolor{mediumorchid}{``fulfill"}$,\textsf{tid},z_{\textsf{tid}}$), which becomes public through HTLC fulfillment\\
			\quad wait until every TU is reported fulfilled by the HTLC interface; otherwise expose its timeout without a sender-success output\\
			\quad set every $\mathsf{State}[\textsf{tuid}_k]\leftarrow(\textsf{tuid}_k,\top)$ and $\mathsf{State}[\textsf{tid}]\leftarrow(\textsf{tid},\top,h_{\textsf{tid}})$\\
			\quad obtain $\sigma_{\textsf{tid}}$ from $\mathcal{G}_{\rm att}$; wait for \textcolor{mediumorchid}{``ok"} from $\textcolor{brickred}{\mathcal{A}}$\\
			\quad send private delayed output $(\sigma_{\textsf{tid}},\textsf{ACK}_{\textsf{tid}})$ to $P_s$ and private settlement acknowledgment to $P_r$
		}
	\end{multicols}
\end{algorithm*}

\textit{(i) Payment initialization:} (Lines \ref{A3l2}-\ref{A3l5}) Before the sender invokes ``\verb|init|", the receiver generates a one-time preimage $z_{\textsf{tid}}$, publishes $h_{\textsf{tid}}=\textsf{H}(z_{\textsf{tid}})$ in the invoice, and retains the preimage. The sender passes the invoice commitment to its assigned smooth node $\textcolor{brickred}{\mathcal{S}_i}$. (Lines \ref{A3l18}-\ref{A3l21}) Before processing payments, $\textcolor{brickred}{\mathcal{S}_i}$ installs $\textsf{prog}$ in $\textcolor{brickred}{\mathcal{G}_{\rm{att}}}$ and receives only $(\textsf{eid}_i,\textsf{mpk}_i)$; the corresponding secret signing key remains internal to $\mathcal{G}_{\rm att}$. (Lines \ref{A3l22}-\ref{A3l24}) During preprocessing, the $\iota$ KMG members, with $\iota\geq2t+1$ and $\tau_{\rm dec}=t+1$, jointly generate a pool of one-time threshold-key entries. Online, they atomically bind one unused entry to the uniformly sampled transaction identifier $\textcolor{colorwheel}{\textsf{tid}}$, return only $(\textsf{pk}_{\textcolor{colorwheel}{\textsf{tid}}},\textsf{vk}_{\textcolor{colorwheel}{\textsf{tid}}})$, and retain their private-key shares inside their enclaves.

\textit{(ii) Payment processing:} (Lines \ref{A3l6}-\ref{A3l12}) Client $\textcolor{brickred}{\mathcal{P}_i}$ encrypts $D_{\textcolor{colorwheel}{\textsf{tid}}}$ under $\textsf{pk}_{\textcolor{colorwheel}{\textsf{tid}}}$, conditionally locks the corresponding funds under $h_{\textsf{tid}}$, and calls ``\verb|pay_t|". Before releasing a share, each KMG member verifies enclave-generated authorization evidence binding the identifier and ciphertext hash and delivers the share only through the attested enclave channel. The enclave invokes $\mathcal{TE}.\textsf{Combine}$ after collecting at least $\tau_{\rm dec}$ valid shares. It splits the recovered demand into $K$ TUs that share $h_{\textsf{tid}}$, allocates one pre-generated key entry per TU, and dispatches the TUs as conditional transfers. The destination enclave applies the same authorization and threshold-decryption checks and returns $\textsf{READY}_{\textsf{tuid}_k}$ only after the corresponding HTLC is irrevocably locked.

\textit{(iii) Complete-set acceptance and acknowledgment:} (Lines \ref{A3l60}-\ref{A3l65}) After all $K$ readiness messages are available, $\textcolor{brickred}{\mathcal{S}_j}$ asks $P_r$ to release $z_{\textsf{tid}}$. It verifies $\textsf{H}(z_{\textsf{tid}})=h_{\textsf{tid}}$ and issues fulfillment for the complete HTLC set with the common preimage. If any TU fails or expires before this point, no preimage is released, the receiver does not accept the payment, and pending transfers follow their timeout branches. After release, individual paths resolve through the underlying HTLC protocol and may complete at different times. Only after every TU fulfillment has been observed does $\textcolor{brickred}{\mathcal{S}_i}$ attest the final state and return $(\sigma_{\textcolor{colorwheel}{\textsf{tid}}},\textsf{ACK}_{\textcolor{colorwheel}{\textsf{tid}}})$ to the sender.

\subsection{The Ideal-World Functionality \texorpdfstring{$\textcolor{brickred}{\mathcal{F}_{SHARE}}$}{F-SHARE}}
Algorithm \ref{F-SHARE} specifies the ideal functionality $\textcolor{brickred}{\mathcal{F}_{SHARE}}$. Each participant $\textcolor{brickred}{\mathcal{P}_i}$ may act as a client or a smooth node, and parties communicate over secure channels. The leakage function $\ell(\cdot)$ explicitly reveals message length and permitted routing metadata but not the protected payload. We use the standard delayed-output convention \cite{Canetti2001}: an output is exposed through the ideal adversarial interface and delivered to its recipient only after acknowledgment. For an encrypted message $m$, that interface receives only $\ell(m)$.

(i) During initialization, the adversary learns the explicitly leaked request metadata, the payment hash, and the transaction identifier. (ii) During processing, it learns $\ell(D_{\textcolor{colorwheel}{\textsf{tid}}})$ and $\ell(D_{\textcolor{colorwheel}{\textsf{tuid}}})$ for each encrypted TU. The functionality records receiver acceptance only after every conditional lock is established and the receiver supplies a valid common preimage; otherwise, it records failure without releasing the preimage. After acceptance, each TU resolves through the underlying HTLC interface, and sender success is returned only after every fulfillment is observed. (iii) The execution proof and payment acknowledgments are delivered privately to their designated callers. This captures black-box execution modulo the stated leakage and the externally visible HTLC outcomes without modeling all paths as one atomic ledger transition.

\subsection{Security Analysis}
We state the scoped security of $\textbf{Prot}_{SHARE}$ in Theorem \ref{theorem1}.

\begin{theorem} \label{theorem1}
	(Scoped security of $\textbf{Prot}_{SHARE}$). Let the KMG contain $\iota\geq2t+1$ members and let $\tau_{\rm dec}=t+1$. Assume that $\Sigma$ is EU-CMA secure, \textsf{H} is preimage and collision resistant, and $\mathcal{TE}$ is $t$-private, robust, and IND-CCA secure when at most $t$ private-key shares are exposed. Assume further that honest KMG members release partial decryptions only after verifying enclave-bound authorization, honest attested programs reveal no information beyond $\ell(\cdot)$ and the stated protocol outputs, and the underlying PCN enforces the stated HTLC and Basic MPP terminal semantics. Against any static PPT adversary that corrupts at most $t$ KMG members before setup, $\textbf{Prot}_{SHARE}$ provides: (i) routing confidentiality for honest requests having identical permitted leakage and public HTLC outcomes; (ii) integrity of accepted enclave execution proofs; and (iii) receiver-side acceptance atomicity for honest endpoints. The third property means that the receiver accepts only a complete HTLC set; it does not assert that all paths resolve simultaneously under arbitrary message scheduling. Relationship unlinkability remains conditional on the underlying PCH construction and is outside this theorem.
\end{theorem}

\begin{proof}
\textit{Routing confidentiality.} Consider two honest transaction or TU payloads with identical permitted leakage, timing metadata, and public HTLC outcomes. The adversary observes the public threshold keys, ciphertexts, authorization evidence, at most $t$ private-key shares, and the values specified by $\ell(\cdot)$. Because $t<\tau_{\rm dec}$, the corrupted shares do not decrypt either challenge ciphertext. Enclave-bound authorization also prevents the adversary from using an honest KMG member or honest enclave as an unauthorized decryption oracle. Replacing the adversary-facing challenge encryptions one at a time therefore yields a standard reduction to the $t$-private IND-CCA security of $\mathcal{TE}$. This comparison does not replace the plaintext consumed by an honest enclave: each execution retains its actual enclave input and internal state, and the theorem compares only executions with the same permitted external behavior. Robustness and share verification ensure that an accepted share set determines one plaintext.

\textit{Execution integrity.} An accepted proof binds the enclave identity, program measurement, command, identifier, ciphertext hash, and resulting state. Producing an accepted proof for an execution not performed by the attested program requires either forging $\Sigma$, forging platform evidence, or rebinding an authorized identifier to another ciphertext with the same hash. These events contradict EU-CMA security, attestation unforgeability, or collision resistance, respectively. Hence an accepted proof corresponds to the stated enclave execution except with negligible probability.

\textit{Receiver-side acceptance atomicity.} Before preimage release, the destination enclave returns $\textsf{READY}$ only for an irrevocably locked TU with sufficient expiry margin. If the complete set is not ready, the receiver withholds $z_{\textsf{tid}}$, accepts no payment, and pending HTLCs remain on their timeout paths. If all $K$ TUs are ready, the receiver releases one preimage whose hash equals the common $h_{\textsf{tid}}$. Under the Basic MPP terminal rule, the honest destination does not fulfill an incomplete set and, once it fulfills any member, issues fulfillment for the complete set. Therefore the honest receiver never accepts a strict subset of the intended payment. After release, the preimage propagates through each route under the ordinary HTLC success and timeout rules; adversarial scheduling may delay individual resolutions, and the sender receives a success acknowledgment only after all TU fulfillments are observed. This proves the scoped acceptance property without treating the $K$ routes as one simultaneous ledger transition.
\end{proof}

\bibliographystyle{IEEEtran}
\normalem % Prevent underlines in bibliography entries.
\bibliography{IEEEabrv, test}

\begin{IEEEbiography}[{\includegraphics[width=1in,height=1.25in,clip,keepaspectratio]{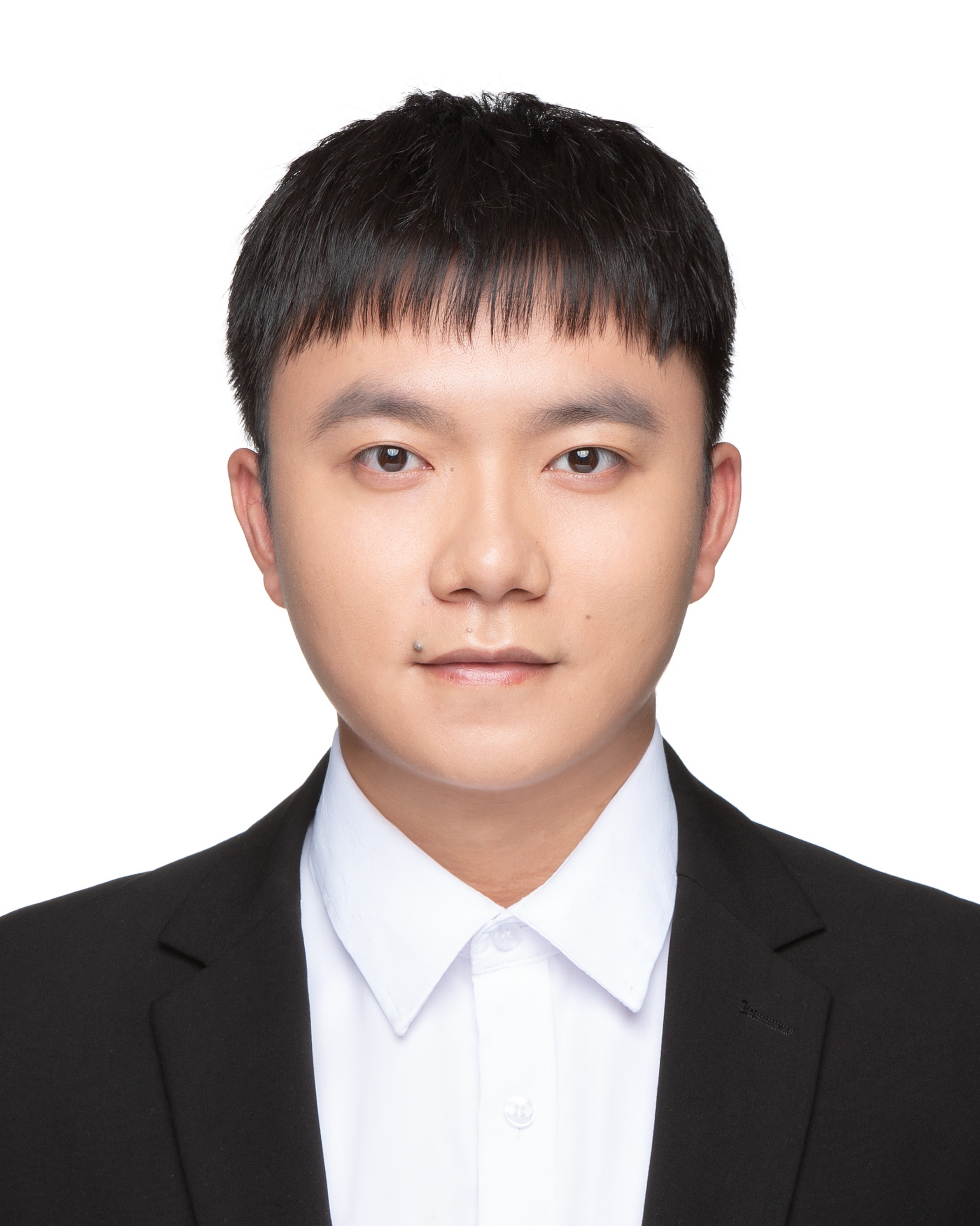}}]{Lingxiao Yang}
	(Member, IEEE) received the B.E. and Ph.D. degrees from Xidian University, Xi'an, China, in 2018 and 2025, respectively. He is currently a Lecturer with the School of Artificial Intelligence and Computer Science, Shaanxi Normal University. His research interests include blockchain technologies and applications, artificial intelligence security, and secure distributed systems. He has published multiple papers in leading IEEE conferences and journals, including IEEE INFOCOM, IEEE Transactions on Services Computing, IEEE Transactions on Computers, and IEEE International Conference on Distributed Computing Systems.
\end{IEEEbiography}

\begin{IEEEbiography}[{\includegraphics[width=1in,height=1.25in,clip,keepaspectratio]{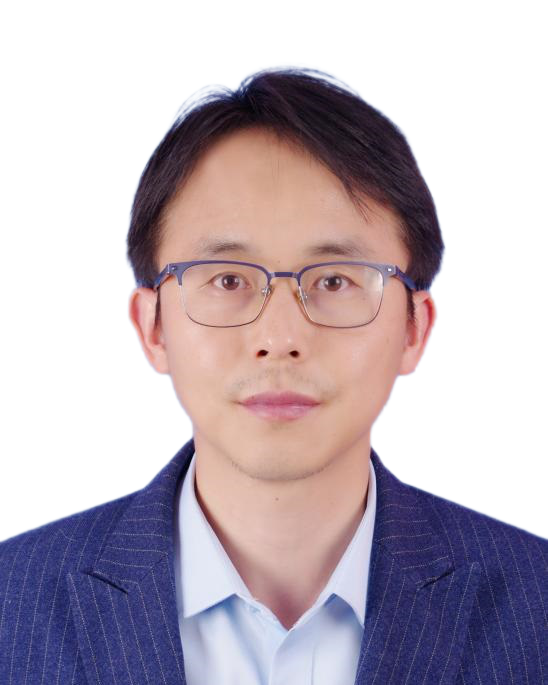}}]{Xuewen Dong}
	(Member, IEEE) received the BE, MS, and PhD degrees in computer science and technology from Xidian University, Xi’an, China, in 2003, 2006, and 2011, respectively. From 2016 to 2017, he was with Oklahoma State University, OK, USA, as a visiting scholar. Now, he is an associate professor with the School of Computer Science, Xidian University. His research interests include cognitive radio networks, wireless network security, and blockchain.
\end{IEEEbiography}

\begin{IEEEbiography}[{\includegraphics[width=1in,height=1.25in,clip,keepaspectratio]{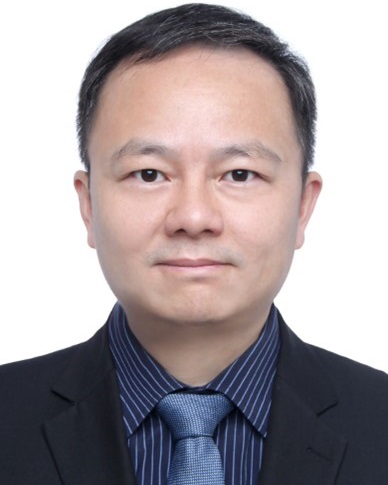}}]{Wei Wang}
	(Senior Member, IEEE) received the Ph.D. degree from Xi’an Jiaotong University, China, in 2006. He is currently a Full Professor with the Ministry of Education Key Laboratory for Intelligent Networks and Network Security, aka MOE KLINNS Laboratory, Xi’an Jiaotong University. He has authored or co-authored more than 100 peer-reviewed papers in various journals and international conferences. His recent research interests include social networks and data security. He is an Editorial Board Member of Computers and Security and a Young AE of Frontiers of Computer Science. He is a Highly Cited Chinese Researcher in Elsevier.
\end{IEEEbiography}

\begin{IEEEbiography}[{\includegraphics[width=1in,height=1.25in,clip,keepaspectratio]{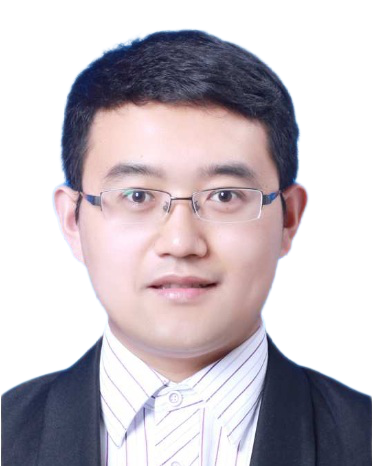}}]{Yong Yu}
	(Fellow, IEEE) received the Ph.D. degree in cryptography from Xidian University, Xi’an, China, in 2008. He is currently a Professor with Shaanxi Normal University, Xi’an. He holds the Prestigious One Hundred Talent Professorship of Shaanxi Province. He has authored over 100 referred journal articles and conference papers. His research interests include blockchain and cloud security.
\end{IEEEbiography}

\begin{IEEEbiography}[{\includegraphics[width=1in,height=1.25in,clip,keepaspectratio]{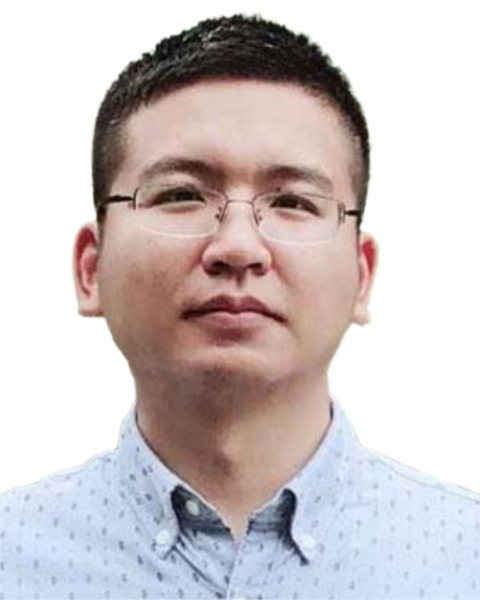}}]{Sheng Gao}
	(Member, IEEE) received the BS degree in information and computation science from the Xi’an University of Posts and Telecommunications, in 2009, and the PhD degree in computer science and technology from Xidian University, in 2014. He is currently a professor with the School of Information, Central University of Finance and Economics. He has authored or coauthored more than 30 articles in refereed international journals and conferences. His research interests include data security, privacy computing, and blockchain technology.
\end{IEEEbiography}	

\begin{IEEEbiography}[{\includegraphics[width=1in,height=1.25in,clip,keepaspectratio]{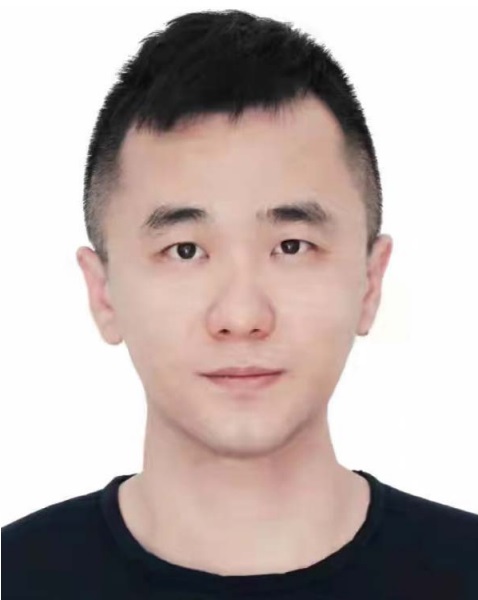}}]{Qiang Qu}
	is a Professor with the University of Chinese Academy of Sciences and the Shenzhen Institute of Advanced Technology, Chinese Academy of Sciences. He is also the Director of the Blockchain Laboratory at Huawei Cloud Tech Co., Ltd. His research interests include blockchain and data-intensive systems.
\end{IEEEbiography}

\begin{IEEEbiography}[{\includegraphics[width=1in,height=1.25in,clip,keepaspectratio]{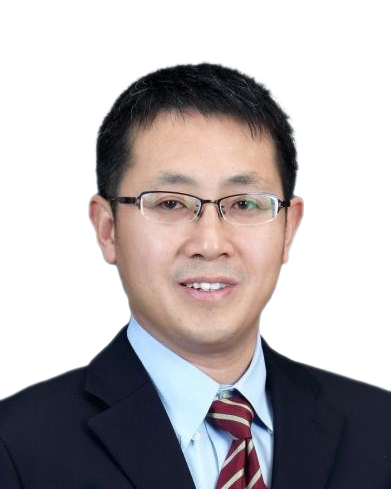}}]{Yulong Shen}
	(Senior Member, IEEE) received the B.S. and M.S. degrees in computer science and the Ph.D. degree in cryptography from Xidian University, Xi’an, China, in 2002, 2005, and 2008, respectively. He is currently a Professor with the School of Computer Science and Technology, Xidian University. He is also an Associate Director of Shaanxi Key Laboratory of Network and System Security and a member of the State Key Laboratory of Integrated Services Networks, Xidian University. His research interests include wireless network security and cloud computing security. He served on the Technical Program Committee of several international conferences, including ICEBE, INCoS, CIS, and SOWN.
\end{IEEEbiography}

\end{document}